\documentclass{aa}
\usepackage[varg]{txfonts}
\usepackage{natbib}
\usepackage[caption=false]{subfig}
\usepackage{longtable}
\usepackage{multicol}
\usepackage{rotating}
\usepackage{pdflscape}
\usepackage{textcomp}

\begin{document}
	
	\title{A new bow-shock source with bipolar morphology in the vicinity of Sgr~A*}
	\author{F. Pei{\ss}ker\inst{\ref{inst1},\ref{inst2}}
		\and M. Zaja\v{c}ek\inst{\ref{inst1},\ref{inst3},\ref{inst6}}
		\and A. Eckart\inst{\ref{inst1},\ref{inst3}} 
		\and N. Sabha\inst{\ref{inst4},\ref{inst1}}
		\and B. Shahzamanian\inst{\ref{inst5},\ref{inst1}}
		\and M. Parsa\inst{\ref{inst1},\ref{inst3}}
		}
	\institute{I.Physikalisches Institut der Universit\"at zu K\"oln, Z\"ulpicher Str. 77, 50937 K\"oln, Germany\label{inst1} \and
		\email{peissker@ph1.uni-koeln.de}\label{inst2}
		\and Max-Plank-Institut f\"ur Radioastronomie, Auf dem H\"ugel 69, 53121 Bonn, Germany\label{inst3}
		\and Center for Theoretical Physics, Polish Academy of Sciences, Al. Lotnikow 32/46, 02-668 Warsaw, Poland\label{inst6}
		\and Institut f\"ur Astro- und Teilchenphysik, Universit\"at Innsbruck, Technikerstr. 25, 6020 Innsbruck, Austria\label{inst4}
		\and Instituto de Astrofísica de Andalucía - CSIC, Glorieta de Astronomía, S/N, 18008 Granada, Spain\label{inst5}
	}
	\date{Received ?? / Accepted ??}

\abstract
{In the direct vicinity of Sgr A* with an approximate projected mean distance of $425\pm 26$ mas we find an extended source. Its sky-projected elongated shape can be described by an averaged spatial extension of $x = 110\pm 20$ mas and $y = 180\pm 20$ mas. With this, the observed object points in the analyzed SINFONI data-sets between 2006 and 2016 directly towards the supermassive black hole. We discuss different possible scenarios that could explain the detected blue-shifted line emission source.}
{Here we present a detailed and extensive analysis of the adaptive optics corrected SINFONI data between 2006 and 2016 with a spatial pixel scale of 0".025 and a corresponding field of view of 0".8 x 0".8 per single data-cube with the focus on the newly discovered source. We spectroscopically identify the source, that we name X8, in the blue-shifted Br$\gamma$ linemaps. Additionally, an upper limit for the continuum magnitude can be derived from the close by S-star S41.}
{For the analysis we applied the standard reduction procedure with the SINFONI/EsoRex pipeline. We applied pre- and post-data correction in order to establish various calibration procedures. For the sharpened images, we are using the Lucy-Richardson algorithm with a low iteration number. For the high-pass filtered images we are using the smooth-subtracting process in order to increase the S/N ratio.}
{We are able to detect the elongated line emission source in quantified data-sets between 2006 and 2016. We find a lower limit for the infrared continuum magnitude of $K_{\rm s} \gtrsim 17.0\pm 0.1$. The alignment of X8 towards Sgr A* can be detected in data-sets that fulfill a sufficient number of observations with a defined quality-level. A more detailed analysis of the results shows indications of a bipolar outflow source which could be associated with either a young stellar object (YSO) or a post-AGB star/young planetary nebula.}
{The near infrared excess source X8 close to S24, S25, and S41 can be detected between 2006 and 2016. Besides an apparent bow-shock morphology, the source shows clear signatures of a bipolar outflow consistent with both a young stellar object and a post-AGB star. If confirmed, this would be the closest bipolar outflow source ever detected close to the SMBH. Similar to the case of the DSO/G2 source and other dusty sources, it further supports the \textit{in situ} star-formation in the direct vicinity of Sgr~A*. If X8 would be a bow-shock source, it would be the third object of this kind that can be found in projection in the mini-cavity. This scenario would support the idea, that the cavity is created by a wind from Sgr A*.}{}

\keywords{stars: chromospheres -- stars: late-type -- stars: winds, outflows -- radio continuum: stars }

\maketitle

\section{Introduction}	
The Galactic center (GC) is a unique laboratory for studying the dynamics of stars and gas in the vicinity of the supermassive black hole (hereafter SMBH) Sgr A*. The nuclear star cluster (NSC) belongs to one of the densest as well as the oldest in the Galaxy and therefore it contains stars of different age and mass \citep[see e.g.][for reviews]{2005bhcm.book.....E,2007gsbh.book.....M,2010RvMP...82.3121G,2014CQGra..31x4007S,eckart2017}. It was argued that the presence of the SMBH, hard radiation, shock waves, considerable magnetic field and a large velocity dispersion inside molecular clouds makes \textit{in situ} star-formation different from the Galactic plane environment \citep{1996ARA&A..34..645M}, which directly affects the initial mass function that is expected to be skewed towards higher stellar masses or have a high low-mass cut-off \citep{1993ApJ...408..496M,2003ApJ...590L..33L}. The discovery of massive and luminous OB stars within $\lesssim 0.5\,{\rm pc}$, which were estimated to have formed within the last 10 Myr, led to the formulation of ``paradox of youth" \citep[][]{Ghez2003}. Subsequently, two main scenarios were suggested to account for the presence of young massive stars -- a stellar cluster inspiral due to a dynamical friction \citep{2001ApJ...546L..39G,2003ApJ...597..312K} and \textit{in situ} star formation in a massive gaseous disc \citep{2004ApJ...604L..45M}; see also \citet{2016LNP...905..205M} for an overview. The cluster inspiral is currently less favoured due to the inspiral timescales that are longer than the estimated age of a few million years of observed OB stars unless the original cluster contained an intermediate-mass black hole \citep{2003ApJ...593L..77H}, but the contribution of both processes is expected over the whole age of the nuclear star cluster. In particular, the current S cluster dynamics is consistent with having originated in an initially disc-like stellar cluster around an IMBH that gets disrupted close to the SMBH, with the IMBH potentially still orbiting the SMBH and influencing S cluster dynamics \citep{2009ApJ...693L..35M}. Additional dynamical processes, such us the disruption of binaries close to Sgr~A* \citep{2003ApJ...592..935G} plausibly contributed to the current dynamical state of the NSC.

In addition, the study of the fast-orbiting, early-B-type star S2 has given new insights into its spectral properties as well as its bound elliptical orbit around Sgr A* \citep[see][]{2017ApJ...847..120H,parsa, gravity2018}. Puzzling NIR-excess dusty sources, such as G1 \citep{witzelG1} or DSO/G2 \citep[see e.g.][]{Gillessen2012,Burkert2012, Eckart2013a, 2014A&A...565A..17Z, Witzel2014, Valencia2015,Zajacek2017} provide a unique opportunity to get a broader understanding of the circumnuclear medium and its dynamics. 

Recently, using high-resolution ALMA observations, \citet{Yusef-Zadeh2017} discovered eleven bipolar-outflow sources in $^{13}$CO, H$30\alpha$, and SiO$(5-4)$ lines that are consistent with having been swept up by jets from young protostars. These outflows exhibit characteristic blue- and red-shifted lobes around their central star with the mass and momentum transfer consistent with protostellar jets from the Galactic disk population. Bipolar-outflow sources in the Galactic centre plausibly manifest a very recent star-formation event that took place within $10^{4}-10^{5}$ years, which is also supported by detecting SiO$(5-4)$ clumps as well as radio-continuum photoevaporative proplyd-like formations within the central parsec \citep{2013ApJ...767L..32Y,2015ApJ...801L..26Y}. In particular, \citet{2013ApJ...767L..32Y} propose that SiO$(5-4)$ clumps manifest high-mass star-formation, while the bipolar-outflow sources analyzed by \citet{Yusef-Zadeh2017} are rather related to low-mass star-formation. There is a general consensus that the \textit{in-situ} star-formation in the Galactic centre region proceeds differently than in the Galactic plane due to the dominant massive point source, which affects the critical Jeans mass due to tidal forces. Two main theoretical models of in-situ star-formation have been developed,
\begin{itemize}
    \item disc-based when a massive gaseous accretion disc surrounding the SMBH fragments into self-gravitating cores, which occurs when the Toomre stability criterion of a disc $Q=c_{\rm s}\Omega/\pi G \Sigma$ is less than unity \citep{1978AcA....28...91P,1987Natur.329..810S,2003ApJ...590L..33L,2004ApJ...604L..45M,2008A&A...477..419C},
    \item cloud-based when an in-falling gaseous clump is prevented from tidal disruption due to the tidal focusing \citet{2013MNRAS.433L..15L,behrang}, radiation or wind pressure or a collision with another cloud, and eventually becomes self-gravitating and fragments into smaller protostellar cores.
\end{itemize}
\citet{behrang} modeled the cloud-based star formation and showed that it can proceed very close to the SMBH thanks to the gravitational focusing of in-falling cold gas, which can be supplied from the circumnuclear disk (CND). The authors show that an initially spherical gas cloud gets tidally stretched due to the presence of the SMBH but at the bf same time, the cloud becomes compressed in the perpendicular direction. This tidal compression can support the formation of compact YSO associations similar to the compact comoving IRS13N association of extremely reddened stars \citep{2008A&A...482..173M}, which is kinematically distinct from the more evolved IRS13E association members \citep{2004A&A...423..155M}. Studying and revealing more potential YSOs near Sgr~A* is highly needed to test and improve the current models of in-situ star-formation in the Galactic centre region.

Despite a large number of observations carried out by various groups with different telescopes of the Galactic center, there are several unresolved issues. 
\begin{figure}[htbp!]
	\centering
	\includegraphics[width=.5\textwidth]{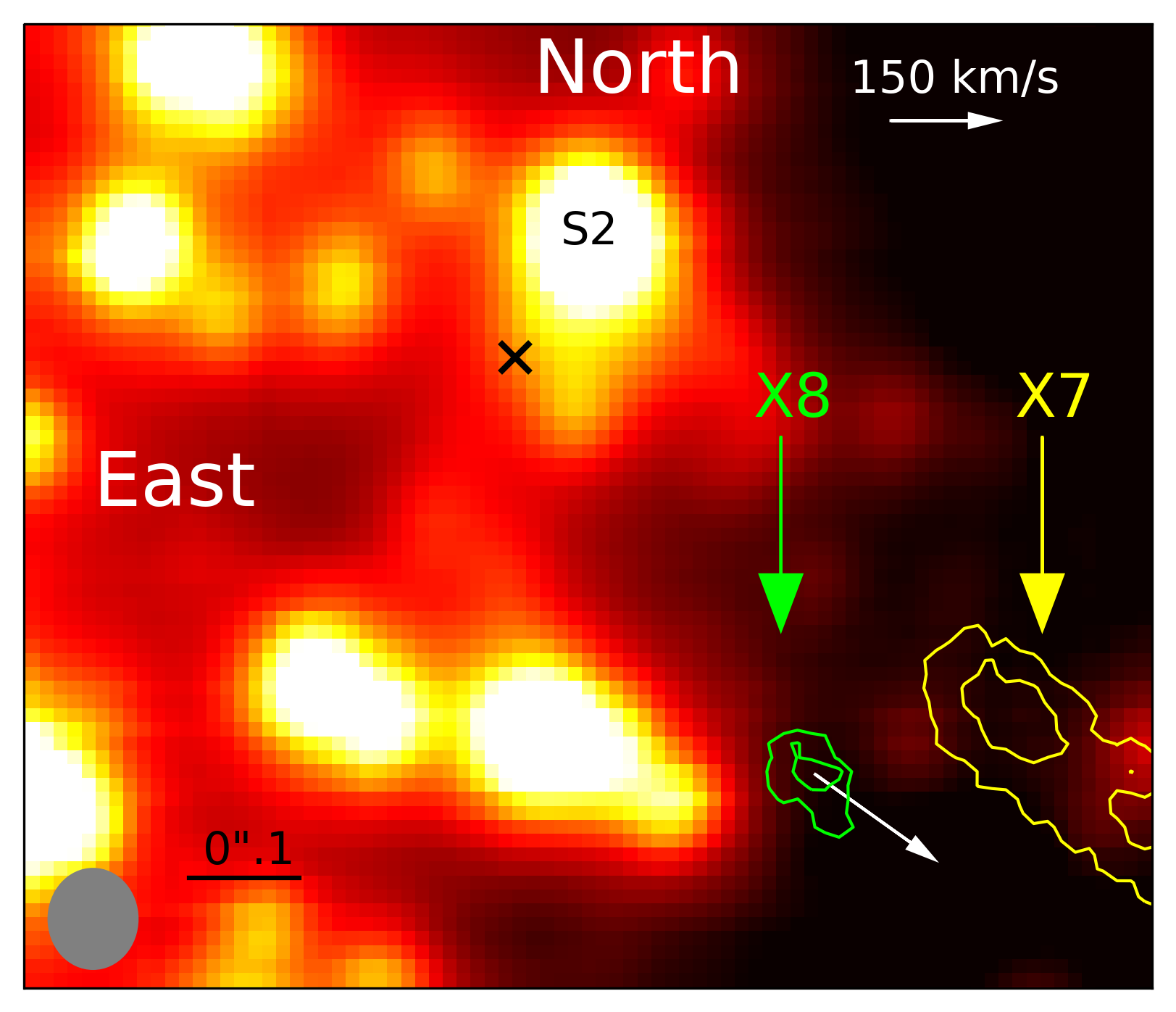}
	\caption{Finding chart of the galactic center. It is a K-band image of the 2015 observations and extracted from the associated H+K SINFONI data-cube. In the lower left corner, the extracted PSF of S2 of the data-cube is given as a grey elliptical patch. The black x next to the B2V star S2 marks the position of Sgr A*. North is up, east is to the left. The length of the black bar in the lower left quadrant is $0".1$ or $100$ mas. The green contour indicates the position of the X8 source while the yellow contour determine the bow-shock object X7. These contours are based on the peak intensity (green line: $62\,\%$ and $84\,\%$, yellow line: $42\,\%$ and $73\,\%$) of the blue-shifted Br$\gamma$ line-maps that are extracted from the H+K SINFONI data-cube. The full size of the bow-shock source X7 cannot be shown since the FOV is limited by the observations.}
\label{finding_chart}
\end{figure}
For example, \cite{muzic2007} discusses a wind/ outflow with the magnitude of $\sim 1000\,{\rm km\,s^{-1}}$ from the direction of Sgr A* that can explain the orientation of comet-shaped sources -- X7 and X3 -- towards the central black hole. These bow-shock sources are moving with a proper velocity of around $300\,km\,s^{-1}$. Based on the observations carried out by NACO, which was mounted at the Unit Telescope (UT) 4, the proper motion of X7 is directed towards the north. In comparison, the proper motion of X3 is directed towards the west. Using the SINFONI instrument mounted on the UT4 at the Very Large Telescope (VLT), we are able to partially track the X7 source between the epochs 2006 and 2016. It is located south-west of Sgr A* at a distance of 400 mas. We observe a blue-shifted line-of-sight (LOS) Br$\gamma$ velocity of around $645\,km\,s^{-1}$ at $\sim 2.1616 \,\mu m$ as well as blue-shifted lines for He$I$ at $\sim 2.0544 \,\mu m$, Br$\delta$ at $\sim 1.9399\,\mu m$, and a weak $Pa\alpha$ line at $\sim 1.8720\,\mu m$. The detection in the Br$\gamma$ line-maps shows an elongated emission that matches the position, alignment, and the described shape of the discovered bow-shock source X7 by \cite{muzic2007}. 
\begin{figure*}[ht!]
	\centering
	\includegraphics[width=1.\textwidth]{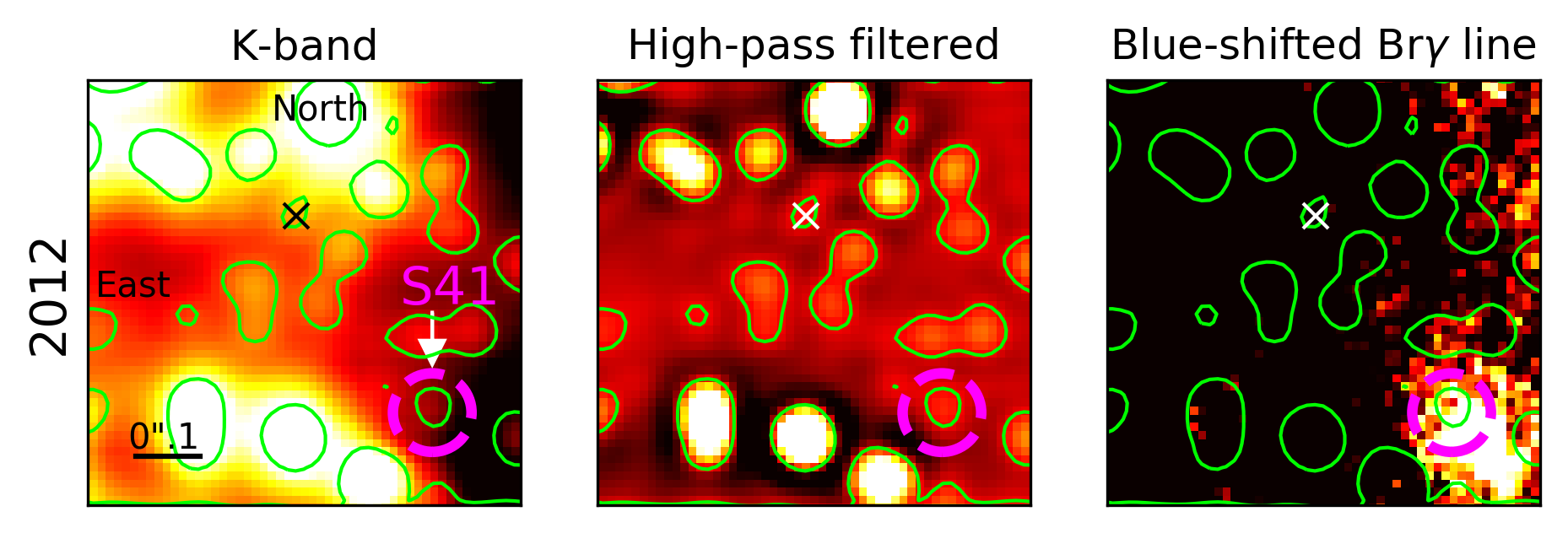} 
	\caption{The Galactic center in 2012 observed with SINFONI with a DIT of 600s per single data-cube. In total, 61 data-cubes are combined with a integrated exposure time of 610 minutes. The resulting on-source exposure time of the final data-cube is around 10 hours. North is up, east is to the left. The lime-coloured contour lines are based on the middle high-pass filtered image. It shows matching positions for the stars in the left K-band image. It is extracted from the SINFONI H+K data-cube. The x marks the position of Sgr A*. The magenta-coloured dashed circle indicates the position of S41. This S-star is in superposition with X8 that is displayed in the right image (bright white object in the lower right corner). This subplot is a blue-shifted line map with respect to the Br$\gamma$ rest wave-length at $2.1661\,\mu m$.}
	\label{2012_contour_map}
\end{figure*}
By applying the same tool-set, which is used to identify X7 in the SINFONI data set, we investigate a new, previously not reported source that we further design as X8. We find a blue-shifted Br$\gamma$ peak at around $2.164 \,\mu m$. Additionally, we report a blue-shifted He$I$ line. Weak signs of a H$_2$ line are also detected. However, a confusion with the Galactic center background emission or the overlapping point spread functions (PSFs) of the S-stars S41 and S24/S25 cannot be ruled out. Because of similarities in shape, alignment, and position between X8 and X7, we discuss the possibility of a bow-shock scenario. In addition, the source appears to possess a bipolar morphology and plausibly falls into the mosaic of young stellar sources as those analysed by \citet{Yusef-Zadeh2017} in the central parsec. The bipolar outflow could be a manifestation of either a young stellar object or alternatively, a late-type star on an asymptotic giant branch and at the beginning of a planetary nebular stage.

Hereafter, we will explain the execution of the observations as well as describe the data reduction process. After Sect. \ref{sec2}, the analysis Sect. \ref{secANA} introduces the used methods like high- and low-pass filtering. The results are presented in Sect. \ref{secRes}. Section \ref{secDis} gives an overview about the different scenarios about the possible nature of X8. Afterwards, we give a conclusion in Sect. \ref{secCon} and summarize our findings.

\section{Observations}
\label{sec2}

\subsection{Data-sets between 2006 and 2016}
The analyzed data between 2006 and 2016 is downloaded from the ESO archive\footnote{\url{http://archive.eso.org/eso/eso_archive_main.html}} with mainly 600 seconds per single data-cube. For every year, the final analysed data-cube is a combination of several single data-cubes. The total on-source exposure time, that is based on these final data-cubes, is given in Tab. \ref{tab1}. 
Additionally, most of the data between 2014 and 2015 of the data was observed with the SINFONI instrument (see \citealt{Eisenhauer2003}) mounted on the VLT/UT4 at Paranal/Chile. 
As in the downloaded data, we used a H+K ($1.45-2.45\,\mu m$) grating but with a 400s exposure time and a spectral resolution of $R\sim 1500$. This translates to a resolving power of $200\, km\,s^{-1}$. We used this exposure time setting to be more flexible in terms of fast changing weather conditions. The pixel scale in every analyzed data-cube is $\bf 12.5\,\times\,25$ mas per pixel, i.e. the smallest available plate scale. Adaptive optics was enabled. The used Natural-Guide-Star (NGS) is located $15".54$ north and $8".85$ east of Sgr A* with a magnitude of around $mag\,=\,14.6$. We rejected the use of a laser guide star (LGS) to avoid the cone effect, since the blue-shifted emission lines of the source could be confused with noise, it is sufficient for the identification to minimize the atmospheric influence. By using a LGS, the possibility of a decreased data quality cannot be ruled out. The Field-Of-View (FOV) is jittered around the position of S2 (see Fig. \ref{finding_chart}) in a way that the B2V star is located close to the upper right quadrant in the fast reduction screen.
During the observations, the airmass has a direct effect on the data quality. If the acquisition of the starting observations is done with airmass values above $1.6$, the AO-loop on the NGS is opened and closed during the observations at a airmass of around $1.1\,\,-\,\,1.2$ to increase the performance of the adaptive-optics correction. 
Comparing the full width at half maximum (FWHM) of S2 before and after this procedure shows a quality increase between $10\%-20\%$. The sky, that is subtracted in the final reduction procedure, is located $5'36"$ north and $12'45"$ west of Sgr A*. It is observed with an exposure time setting of 400 s in the H+K domain. The science template (o) and the sky template (s) of the SINFONI Observation Block (OB) is arranged in a o-s-o pattern. This arrangement ensures the highest data-quality combined with an efficient observation-time usage. It also minimizes the negative influence on the line shape of weak signals close to the noise level that is caused by the variation of the sky emission lines (see \citealt{Davies2007}).
\begin{figure*}[htbp]
	\centering
	\includegraphics[width=1.\textwidth]{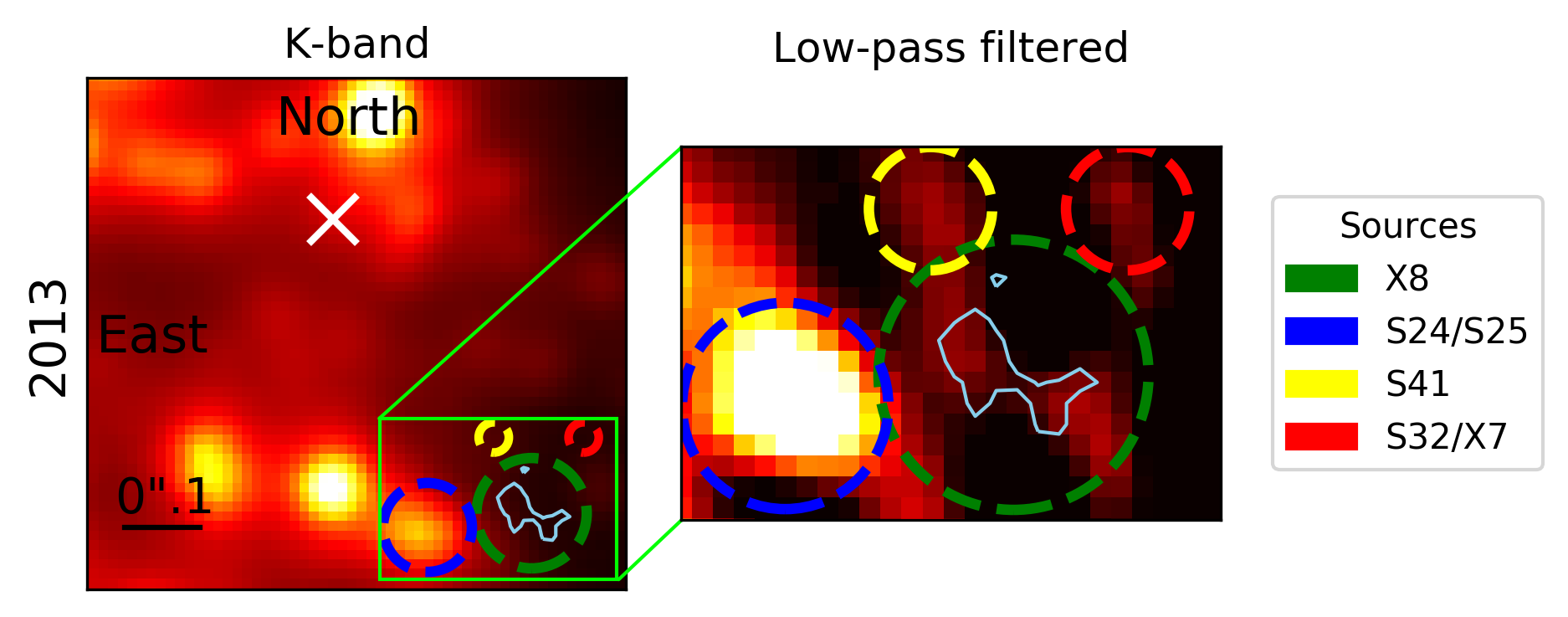} 
	\caption{The Galactic center in 2013. The on-source observation time is almost 30 hours. Here we extracted a K-band image from the final data-cube. On the right side, a zoomed in and low-pass filtered image of the green rectangle area is presented. The green dashed circle shows the position of X8. The contour line is based on the blue-shifted Br$\gamma$ line map of 2013.}
	\label{2013_contour_zoom}
\end{figure*}

\subsection{Data-set of 2005}
For the sake of completeness it should be mentioned that also data from 2005 is available in the ESO archive. Unfortunately, this data set does not provide a sufficient number of exposures that cover the area of X8. This leads to a unsatisfying detection of X8 that could also be interpreted as random artifacts which will be briefly discussed in the following.
As mentioned before, the low number of single data-cubes, which we stack to create the final cube, could cause a shifting centroid position since the on-source time is not enough to suppress low frequency peaks caused by the noise. Therefore, the wings of the investigated Br$\gamma$ line are more broadened and dominate the emission peak itself. On the other hand, the spectral difference of the line peak between 2005 and 2006 is around $0.000275\,\mu m$ or $\sim 38\,km\,s^{-1}$. With a resolution in the H+K grating of $R=1640$, we calculate a spectral resolution of $0.0013\,\mu m$ for the peak in 2006. In 2005, we are using K-band data with a difference between 2005 and 2006 of $0.00014$. The K-band resolution is increased to $R=4490$. That results in a spectral resolution of $0.00048\,\mu m$. We can therefore conclude that the fluctuations are the result of the limited resolution of SINFONI and the low number of observation nights. It is worth noting that we can also observe these fluctuations over the emission area of X8 by using a single pixel aperture. Some pixels show a somewhat smaller velocity compared to the majority of the detection. Even though this can be traced back to the discussed scenarios. It could also indicate the presence of an outflow, a bow-shock feature, or a more complex structure in general.

\subsection{Data reduction}

We use the standard procedure (ESO pipeline with the GUI Gasgano) for reducing the SINFONI data (see \cite{Modigliani2007}). In order to classify the quality of the data, we use the FWHM of S2. The best result can be achieved with a FWHM of less than 6 pixel in x- and y-direction for the single data-cubes. With a pixel scale of 12.5 mas, this threshold translates to 75 mas in x- and y-direction.  However, values up to 7.5 pixel (93 mas) for the FWHM can also provide good data quality. This originates in the background emission of the galactic center as well as overlapping FWHM of the surrounding stars. A case-to-case study is needed to identify data-cubes with a sufficient S/N ratio. For the instrument setup, Dark, Linearity, Distortion, and Wave frames are provided by ESO. After the reduction, we establish various data correction procedures. For that, we are using DPuser (\cite{Eckart1991}, and T. Ott) and IDL routines. Firstly, we remove bad pixels that are located mostly at the border of the data-cubes. This is important for the stacking of the individual cubes that takes place at the end of the correction process. The result would be reflected in non-continuous pixel values. Secondly, we correct for the pixel to pixel variations on the detector by multiplying single rows with a proper factor. After that, the telluric absorption is corrected by taking several aperture star spectra from data-cubes that are combined by individual days. The IDL routine applies a polynomial fit to every spectra and corrects afterwards the telluric absorption.
In the last step, we stack the already shifted and normalized data-cubes according to a reference frame that is created with single K-band images of the GC. The K-band images are extracted from single SINFONI data-cubes by selecting the wavelength range between $2.0\,\mu m\, -\,2.2\,\mu m$. By choosing sufficient cubes, a continuum reference frame with $1".5\times \, 1".5$ is created.

\section{Analysis}
\label{secANA}

For the analysis we used different techniques that we introduce in the following subsections.

\begin{table*}[htbp!]
 								\centering
 								
 								\begin{tabular}{ccccccccc}
 									\hline\hline
 									\\		Date & \multicolumn{3}{c}{Distance to Sgr A*} & Position angle &  \multicolumn{2}{c}{Spatial extension}& On-source exp. &Total exp. time \\  & \cline{1-3} \cline{6-7} & Total & $\Delta$RA & $\Delta$DEC &  &  x-direction & y-direction & & \\
 									(YYYY) & (mas) & (mas) & (mas) & (°)  & (mas) & (mas) &   & (min) \\ \hline 
 									
 							\\	2006 & 380  & 210 &  310 & 45.5 &  130 & 230 & 24  & 240  \\
 								2007 & -    & -   &  -   &  -   &   -  &   - & 17  & 170  \\
 								2008 & -    & -   &  -   &  -   &   -  &   - & 21  &  210 \\
 								2009 & 390  & 210 &  330 & 67.5 &  120 & 170 & 27  &  270 \\
 								2010 & 430  & 230 &  360 & 41.7 &  140 & 200 & 64  &  640 \\
 								2011 & 440  & 250 &  360 & 47.3 &  110 & 180 & 43  &  430 \\
 								2012 & 440  & 230 &  370 & 59.7 &  120 & 170 & 61  &  610 \\
 								2013 & 450  & 250 &  370 & 48.0 &  080 & 190 & 176 &  1760 \\
 								2014 & 480  & 260 &  410 & 52.3 &  110 & 170 & 218 &  1454 \\
 								2015 & 470  & 260 &  380 & 54.2 &  080 & 140 & 49  &  327 \\
 								2016 & 500  & 300 &  400 & 54.8 &  130 & 190 & 54  &  360 \\						
 									\hline	\\
 								\end{tabular}
                                \caption[X8 analysis]{The properties of X8 between 2006 and 2016. The position-error can be determined to $\pm 1.5$ pixel that can be translated into $\pm 20$ mas for every position value given for $\Delta$RA and $\Delta$DEC. The error of the total distance of X8 to SgrA* results with error propagation in $\pm 20$ mas. In every year, a 8 pixel Gaussian is fitted to the blue-shifted Br$\gamma$ line-maps of X8. In the data-cubes that cover the years 2007 and 2008, X8 is located at the border of the FOV. Parts of the emission are not covered by these data-cubes. The spatial extension is referring to the long (y-direction) and short axis (x-direction) of X8. The position angle (PA) is measured east of north.}							
                                \label{tab1}
\end{table*}

\subsection{High-pass filtering}

Since overlapping PSFs as well as noise can complicate the identification of the X8 source, we use high-pass filtering to decrease the chance of confusion and to identify single objects that are close to the detection limit. Not only can the quality of the data influence the detection of X8, also the close distance to S24, S25, and S41 plays a major role in the analysis of the source. Figure \ref{2012_contour_map} shows the blue-shifted Br$\gamma$ line map (right subplot). The figure shows, that X8 is partially blocked by the presence of the S-star S41. Therefore, High-pass filtering cannot be used to identify X8 in the continuum. However, it can be used to determine a upper limit for the magnitude. The PSF of this star decreases the chance of a successful identification in the continuum. For the smooth-subtracted image (middle subplot) we use the filtering technique that is described in \cite{muzic2007}. The authors of the mentioned publication subtract a Gaussian smoothed version of the original continuum image and smooth the output again. With that, high frequencies pass while low frequencies, like noise, are reduced. Because of different weather conditions and consequently different PSFs in the years between 2006 and 2016, we adjust the Gaussian to 3-5 pixel for the subtracted and re-smoothed continuum images. 
\subsection{Low-pass filtering}

The crowded FOV in the galactic center demands additional filtering techniques. The iterative Lucy-Richardson deconvolution algorithm \citep{Lucy1974} can be used to sharpen the image. For that, we used 20 iterations and analyzed the convolved image without deconvolving the result. The low number of iterations preserved the positions of the stars in the field of view and ensures that no artificial sources are created. Therefore, analyzing the convolved image is justified. We extracted a K-band image from the final data-cube by selecting the spectral range between $2.0\, \mu m\, - \, 2.2\,\mu m$. Since the final data-cube is a stacked result of many single observations that are observed under different weather conditions, an artificially created PSF can reflect this circumstance with satisfying results. The only star that can be used to extract a natural PSF in the GC with SINFONI in a FOV of $1".0\,\times\, 1".0$ around Sgr A* is S2. However, some stars like S13, S56, and S64 are in superposition with S2. Therefore, a natural PSF will not lead to representative results for the GC region. An artificial PSF (APSF) neglects the influence of overlapping PSFs. It can be created by applying a Gaussian filter to an empty array. The size needs to match the dimensions of the analyzed data. For the extracted K-band image, we are using an APSF that is rotated by 45°(counterclockwise) with a spatial extension of $4.0$ pixel in $x$- and $4.5$ pixel in $y$-direction. The size of K-band image as well as the APSF array is increased to $256\,\times\,256$ pixel. S2 is centered at $(128,128)$, the same centering is applied to the APSF. After the setup of the normalized files, the Lucy-Richardson algorithm (QFitsView, T. Ott) is used to sharpen the image.  

\subsection{Line and continuum maps}
The line-maps represent the blue-shifted Br$\gamma$ emission at $2.164\,\mu m$. Compared to the blue-shifted Br$\gamma$ line, the He$I$ emission line is characterized by a decreased intensity of $\sim 50\%$. The more prominent blue-shifted Pa$\alpha$ line shows an increased level of confusion due to the telluric absorption lines. This line is therefore excluded and will not be discussed in the further analysis. The Br$\gamma$ line-maps are achieved by subtracting the K-band continuum from the emission peak itself. In addition, we stack Br$\gamma$ line-maps of several years that fulfill the needed requirements to the FOV and data quality. This is done by shifting the individual line-maps of X8 to a fixed position. We used the center of gravity of the $50\%$ contour line to determine the exact position of X8. The white inset located in the upper left quadrant in the image shows a cut along the alignment axis which is directed towards Sgr~A*.
For the high-pass filtered image (Fig. \ref{2012_contour_map}, middle), a continuum frame from the final combined data-cube is used. We select the spectral range of $2.0\,\mu m$ - $2.2\,\mu m$ in the SINFONI data-cube and extract the K-band continuum image. 
After that, the image is normalized.

\section{Results}
\label{secRes}
In this section we apply the discussed methods and tools of section \ref{secANA} to the discovered X8 source. With these tools, we are also able to detect the confirmed bow-shock source X7 in selected data sets.
\begin{figure*}[htbp!]
	\centering
	\includegraphics[width=1.0\textwidth]{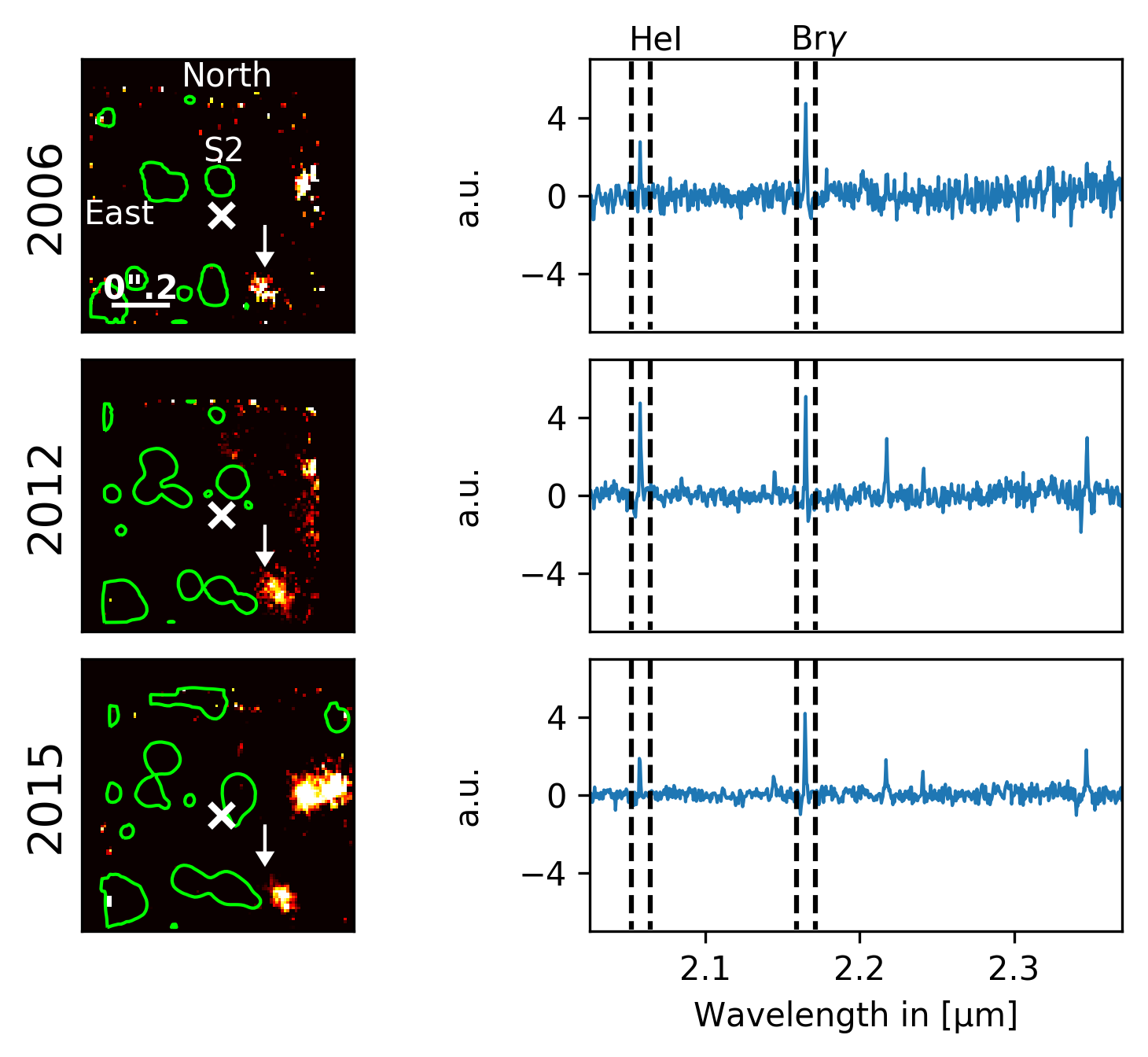}
	\caption{The blue-shifted Br$\gamma$ line-maps of the Galactic center at $2.164\,\mu m$ (left) that are based on the data sets of 2006, 2012, and 2015. The green contour lines are extracted from the K-band continuum emission of the associated data-cubes. Sgr A* is located at the x, the S-star S2 is labeled. The arrow points towards the position of X8 in the blue-shifted Br$\gamma$ line-maps in 2006, 2012, and 2015. The x is centered on SgrA* in every year. The arrow is fixed on the position of X8 in 2006 to show the proper motion. On the right side, the K-band spectrum between $2.0\,\mu m$ - $2.4\,\mu m$ of X8 is presented. The intensity is given in arbitrary units (a.u.) and the blue-shifted He$I$ and Br$\gamma$ line are framed with a dashed line. The spectrum of 2012 and 2015 shows blue-shifted [Fe III] multiplet ${}^3 G\,-\,{}^3 H$ at $2.1410\,\mu m$, $2.2142\,\mu m$, $2.2379\,\mu m$, and $2.3436\,\mu m$.}
	\label{x8_overview}
\end{figure*}
\subsection{X7 detection}
Based on the presented analyzing tools in Section~\ref{secANA}, we identified the bow-shock source X7 \citep[see][]{muzic2007, muzic2010} in the available SINFONI data-set between 2006-2016. The investigated line-maps of the data-cubes show a blue-shifted Br$\gamma$ line at $\sim 2.1614 \,\mu m$ with a related velocity of around $645\,km\,s^{-1}$. The contour of the X7 line map is shown in Fig. \ref{finding_chart}. The spectral analysis shows also blue-shifted lines for He$I$ at $\sim 2.0544 \,\mu m$ $(\sim 625\,km\,s^{-1})$, Br$\delta$ at $\sim 1.9399\,\mu m$ $(\sim 800\,km\,s^{-1})$, and a Pa$\alpha \,\mu m$ line at $\sim 1.8720\,\mu m$ $(\sim 577\,km\,s^{-1})$. We also detected a CO line at $2.3226\,\mu m$ in the spectrum of X7. However, the line-map analysis does not show an emission counterpart. Since this wavelength corresponds to the CO rest-wavelength, we can conclude that this line likely belongs to the background or foreground emission.\newline
With a spectral uncertainty of around 100 km/s and the non-negligible influence of telluric absorption lines, the additional velocities are in agreement with the detected blue-shifted Br$\gamma$ line speed. The K-band continuum is dominated by background emission and stars that are close to X7. Especially S32 (see Fig. \ref{2013_contour_zoom}) is in superposition with the K-band continuum emission of X7 and contributes to the confusion of detecting the bow-shock source. \cite{muzic2010} derive a K-band magnitude for X7 of mag = $16.9 \pm 0.1$. For the K-band magnitude of the star S32, \cite{Gillessen2009} derives a magnitude of mag = $16.6$. With the reference star S2, we calculate a magnitude for S32 of mag = $17.06 \pm 0.06$ based on the data-cube of 2013. This results in an averaged magnitude of mag = $16.83\pm 0.06$. These mixed results are reflected by the circumstance that X7 and S32 coincide in position and magnitude. This, however, leads to unsatisfactory results when using the smooth-subtract algorithm in order to identify X7 in our K-band continuum data. With respect to X7, X8 exhibits comparable properties (position, magnitude, composition, alignment, LOS velocity). In this picture, X7 and X8 are likely of the same nature and could possible be of the same origin. Hence, X8 seems to be the third observed bow-shock source that can be found in projection in the the mini-cavity which is supposed to be created by a wind/outflow emanating from the direction of Sgr A*.

\begin{figure*}[htbp]
	\centering
	\includegraphics[width=0.8\textwidth]{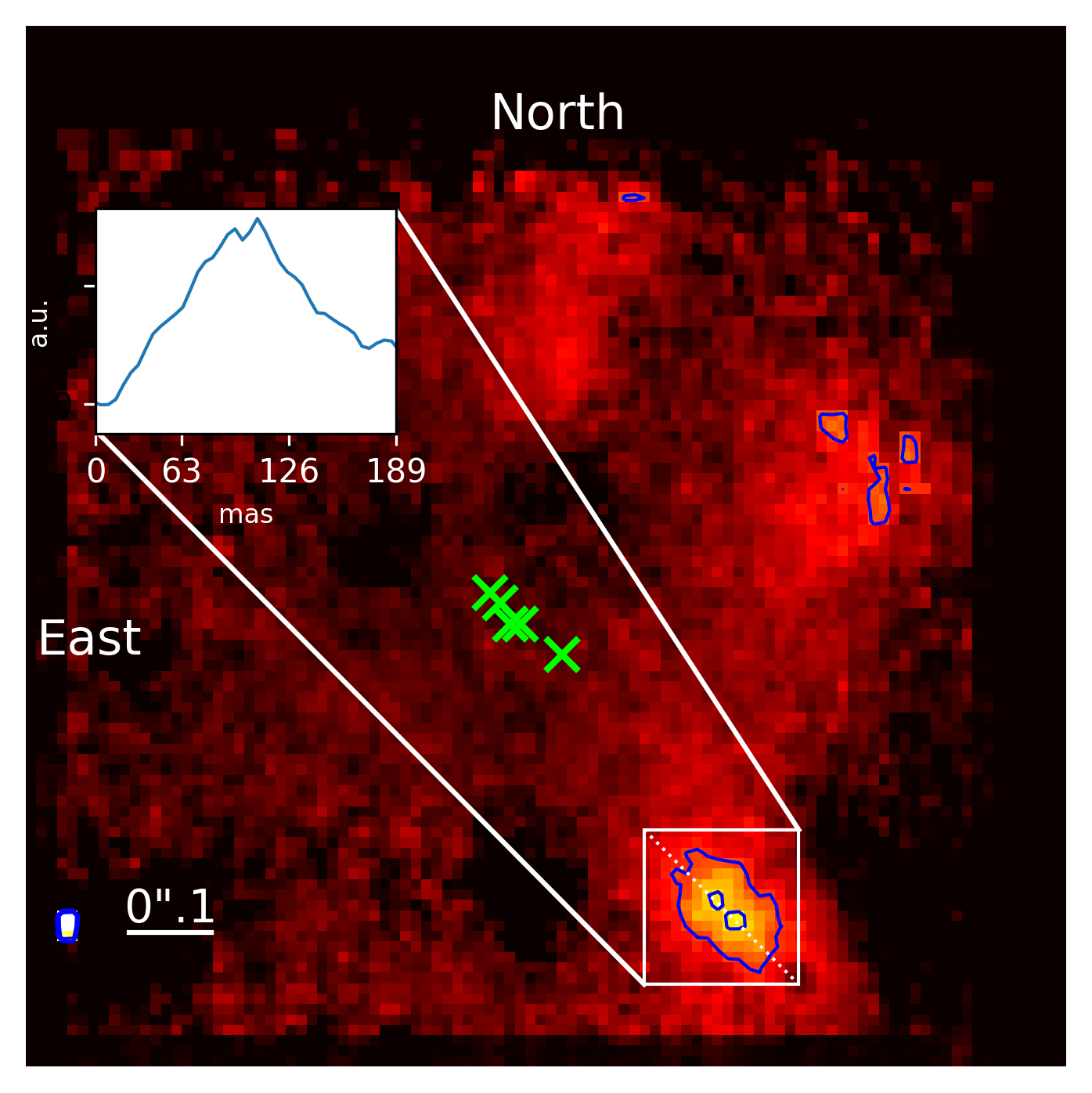} 
	\caption{The stacked Br$\gamma$ line map showing X8. The image is based on the 2006, 2012, 2013, 2015, and 2016 data-set. The data sets of 2008, 2009, 2010, 2011, and 2014 are partially covering the X8 emission area or show effects from a reduced S/N ratio. Since non-continuous effects can be found at the border of the data-cube, the data-sets of 2008, 2009, 2010, 2011, and 2014 are excluded from the stacking. A reduced S/N ratio results from a low number of exposures covering the X8 emission area or a non satisfying data-cube correction. Because the single Br$\gamma$ line-maps of 2006, 2012, 2013, 2015, and 2016 are centered on X8, the stacked figure implies an apparent movement of Sgr A* that is indicated by a green x. The y-axis of the white patch in the upper left corner is in arbitrary units (a.u.).}
	\label{x8_stacked}
\end{figure*}
\begin{figure*}[htbp!]
	\centering
	\includegraphics[width=0.8\textwidth]{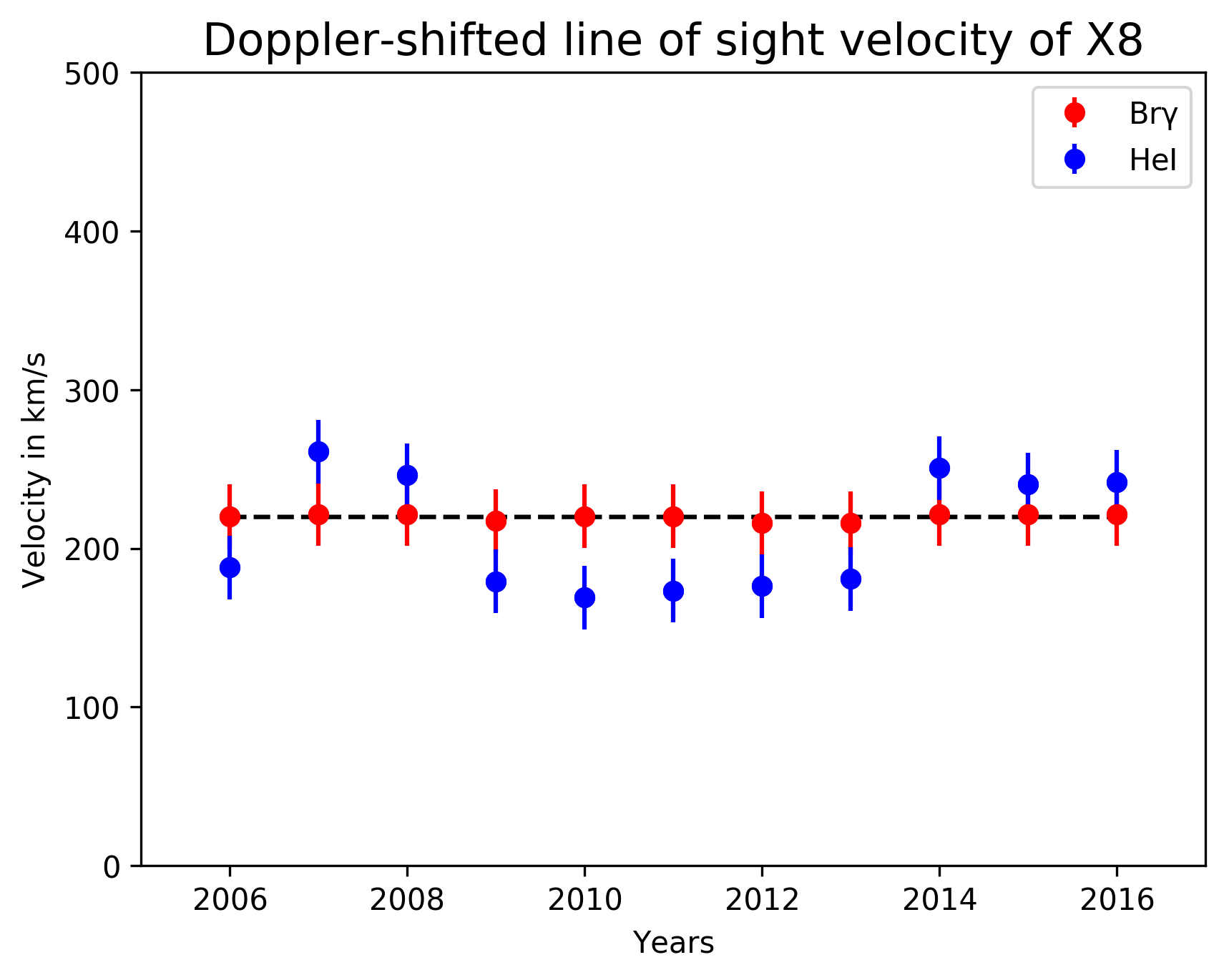}
	\caption{The LOS of X8 between 2006 and 2016. It is based on the blue-shifted centroid Br$\gamma$ (red data-points) and He$I$ (blue data-points) velocity of X8 with an error of $\pm 20\,km\,s^{-1}$ for every data-point. The dashed line represents a linear fit to the blue-shifted Br$\gamma$ data-points.}
	\label{x8_velocity}
\end{figure*}

\subsection{X8 detection}
In Fig. \ref{x8_overview} we present a selection of the X8 detection in the SINFONI data covering 3 years (2006, 2012, and 2015). The first column shows a blue-shifted Br$\gamma$ line map with continuum contours that are extracted from the final data-cubes. The position of Sgr A* is marked with a white $x$. We centered every line map on the position of Sgr A* at (44,47) in the $100\,\times\,100$ pixel array. The B2V star S2 is indicated by its name, while a white arrow (fixed position in every presented line map) shows the introduced X8 source. Based on the derived position of Sgr A*, X8 moves with a proper motion of around $300\,km\,s^{-1}$ away from SMBH. X8 can be found in every year in a comparable area. It is slightly elongated with a mean FWHM of $x=110$ mas and $y=180$ mas. That indicates an extended shape because of the SINFONI PSF of around $60\,-\,80$ mas in $x$- and $y$-direction. In the presented data set, we also detect hints of the D2 and D3 dust sources (\citealt{EckartAA2013}) at the upper right border of the FOV.
The second column of Fig.~\ref{x8_overview} shows the related spectroscopic K-band emission lines of X8 that are extracted with a 5-pixel circular shaped aperture from the final data-cubes. We find blue-shifted lines like Br$\gamma$ at $2.164\,\mu m$ and He$I$ at $2.057\,\mu m$ with a FWHM of around $170\,km\,s^{-1}$. These wavelengths correspond to an averaged blue-shifted velocity of $\sim\, 200\,km\,s^{-1}$ with respect to their corresponding rest wavelengths at $2.1661\,\mu m$ (Br$\gamma$) and $2.0586\,\mu m$ (He$I$). Additionally, we find two other lines at $2.2176\,\mu m$ and $2.3471\,\mu m$. It is not clear to which rest-wavelength these lines can be linked since either the resulting velocity does not match the blue-shifted Br$\gamma$ and He$I$ lines or the lines match the velocity but are red-shifted instead. These emission lines are weaker compared to the detected Br$\gamma$ and He$I$ by a factor of 2-4 and not free of confusion. Mainly the background and the surrounding stars influence the detection of X8. But also the data quality, the amount of data, and the weather conditions play a non-negligible role in the X8 analysis. It could be speculated that these lines indicate a more complex structure or possible outflow properties. For Fig. \ref{x8_stacked} we are using the individual blue-shifted Br$\gamma$ line-maps. We select data-cubes that match our requirements in terms of the field of view and the data quality. Then we apply a shifting vector to the line-maps such that X8 is at the same position in every used data-set. At the expected position of X8, the stacked result shows an elongated object that is pointing towards Sgr A*. It shows two central lobes that are highlighted with blue contour lines at $60\%$ and $90\%$. The white inset is a 2D cut along the alignment of X8 and reflects the finding of the two emission peaks that are around $12.5\,-\,25$ mas apart. The two lobes are embedded in the elongated and blue-shifted Br$\gamma$ source. The object itself is directed towards Sgr A* that is represented here with a green $x$ for every single year. Since Fig. \ref{x8_stacked} is a stacked image of several years of a moving source, Sgr A* seems to move. Of course, the X8 source moves and the shift of Sgr~A* reflects its motion since the reference frame is centered on it. The Br$\gamma$ emission sources D2 and D3 can be partially found at the right border in the upper FOV because the blue-shifted wavelength of the two sources is close to the X8 emission. Also, another faint emission source is visible around 0".1 in the $y$-direction above Sgr A*.

\subsection{Proper motion and line-of-sight velocity}

For determining the proper motion, we infer the offset between X8 and Sgr~A* in the blue-shifted Br$\gamma$ line map in 2006 ($386$ mas) and 2015 ($462$ mas). In order to pinpoint the exact distance, we use the center of gravity of the 50 percent contour line in the channel maps. These measurements result in $\Delta d=70$ mas for X8 between 2006 and 2015. With a distance to the GC of $8\,kpc$ and the time difference between 2006 and 2015, we get a proper motion of $v=293\,\pm 20\,km\,s^{-1}$. 
For the line-of-sight velocity (Fig. \ref{x8_velocity}), we are using the blue-shifted He$I$ and Br$\gamma$ centroid line. The blue-shifted He$I$ line as well as the blue-shifted Br$\gamma$ line detection is free of confusion because they are spectrally isolated. However, we derive a mean He$I$ velocity of $v=210\,\pm 20\,km\,s^{-1}$ whereas the slightly increased Br$\gamma$ velocity is $v=220\,\pm 20km\,s^{-1}$. These variations between the blue-shifted He$I$ and Br$\gamma$ line detection are based on the spectral resolution as well as instrumental effects that are described in \cite{Eisenhauer2003}.
It shows, that the line-of-sight velocity is constant within uncertainties for all used epochs. This is also reflected in the presented spectrum of X8 (see Fig. \ref{x8_overview}). The average velocity is around $215 \,km\,s^{-1}$ and therefore close to the detected proper motion of around $v=290\,km\,s^{-1}$. 

\begin{figure*}[htbp]
	\centering
	\includegraphics[width=1.\textwidth]{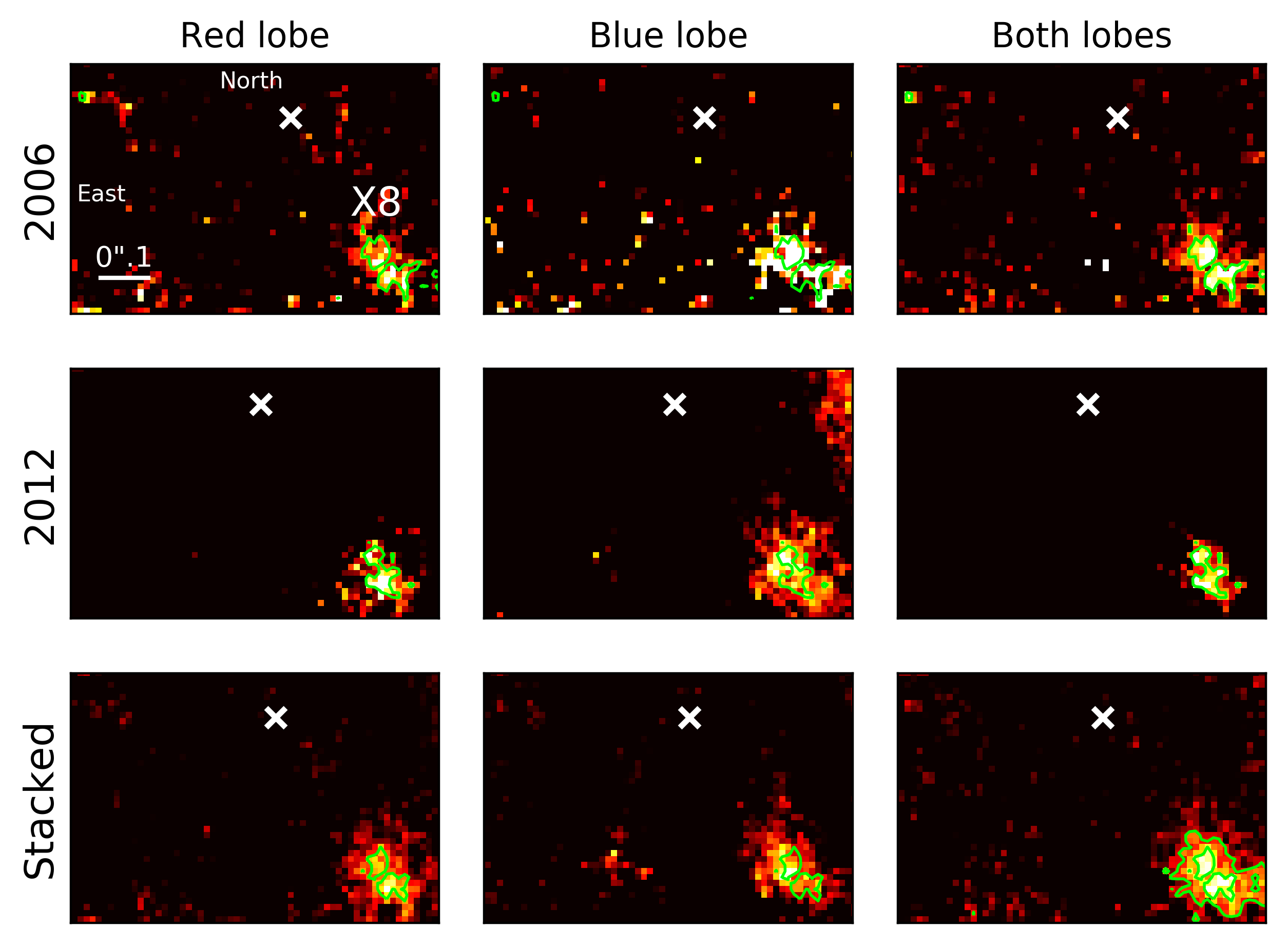}
	\caption{Blue-shifted Br$\gamma$ line-maps of X8 in 2006 and 2012. A scale valid for all images is given in the top left panel. The last row shows the stacked result of 2006 and 2012 with an averaged position of Sgr A* (white x) that is based on the presented years. The red lobe column represents the wavelength range of $2.16455\,\mu m\, - \, 2.16516\,\mu m$. The blue lobe results cover a wavelength range of $2.16394\,\mu m\, - \, 2.16455\,\mu m$. For the last column, we select the wavelength range of $2.16394\,\mu m\, - \, 2.16516\,\mu m$ that results in the presence of both lobes.}
	\label{x8_line}
\end{figure*}
\subsection{Velocity gradient}

As indicated in Fig. \ref{x8_stacked}, the blue-shifted source X8 contains two central lobes. With a constant FWHM of around $170\,km\,s^{-1}$, we are able to select the red- and blue-wing of the Br$\gamma$ emission line. The faster blue-shifted wing with respect to the Br$\gamma$ rest-wavelength covers the wavelength range between $2.16394\mu m\, - \, 2.16455\mu m$. The slower red-shifted wing of the emission line covers a range of $2.16455\mu m\, - \, 2.16516\mu m$. With that, we detect a velocity gradient over X8 that is reflected by an variation of the emission towards a certain area of the blue-shifted Br$\gamma$ source as a function of the selected wavelength range. The mean velocity of the velocity-gradient slope is around $215\,km\,s^{-1}$. The mean velocity of the blue-shifted wing is $177\,\pm\, 38\,km\,s^{-1}$ whereas the mean velocity of the red-shifted wing is $257\,\pm\, 42\,km\,s^{-1}$. With that, the velocity gradient can be determined by $45\,\pm\, 10\,km\,s^{-1}\, / \, 12.5\,mas$.
\begin{figure*}[htbp]
	\centering
	\includegraphics[width=1.\textwidth]{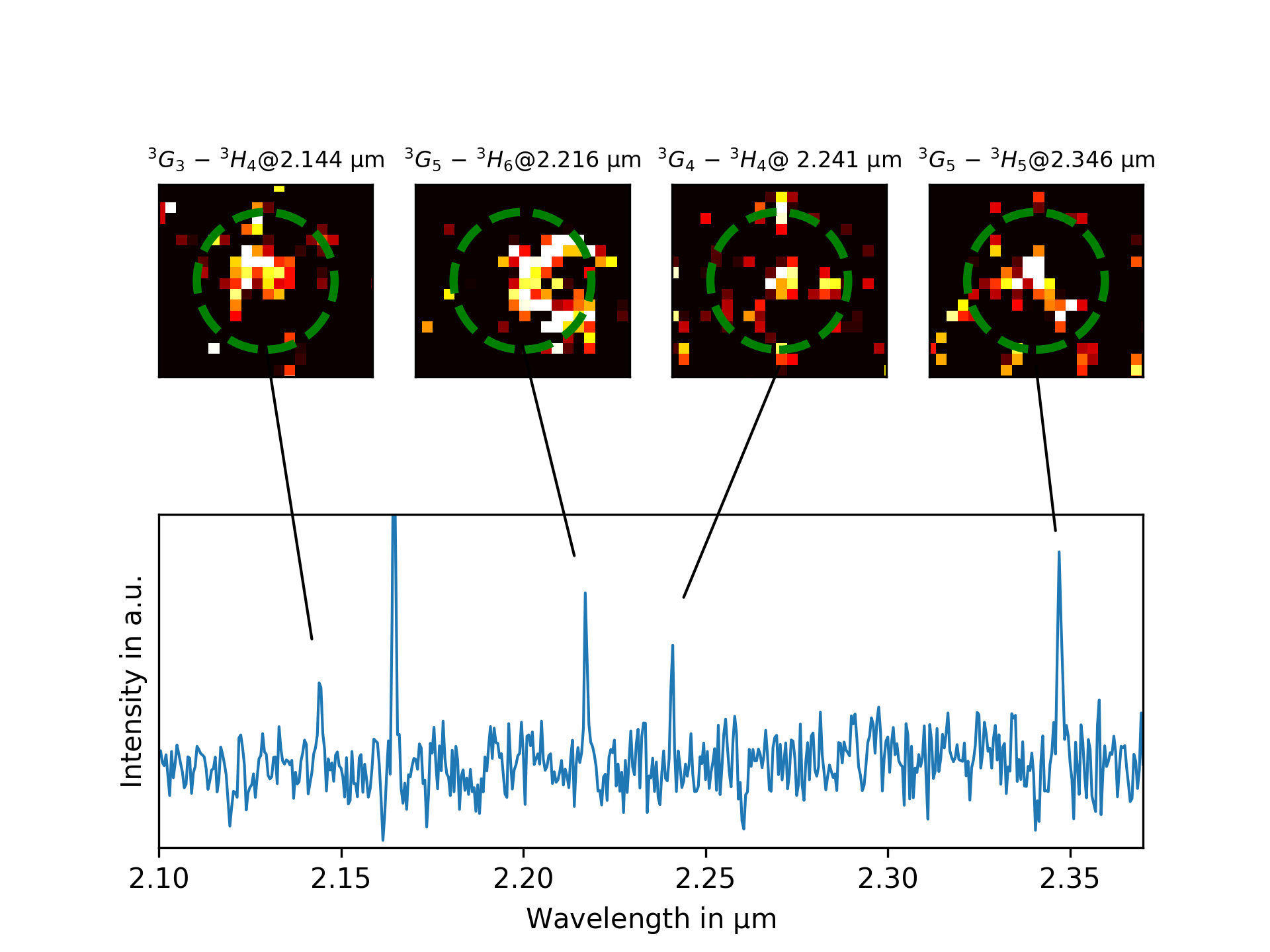}
	\caption{Blue-shifted [Fe III] line maps of X8 in 2015. The spectrum shows the [Fe III] multiplet. At $2.164\,\mu m$, the prominent blue-shifted Br$\gamma$ is located. The four line-map are a $19\times 19$ px ($230\times 230$ mas) cutout from the data-cube of 2015.}
	\label{X8_Fe_line}
\end{figure*}
\begin{figure*}[htbp]
	\centering
	\includegraphics[width=1.\textwidth]{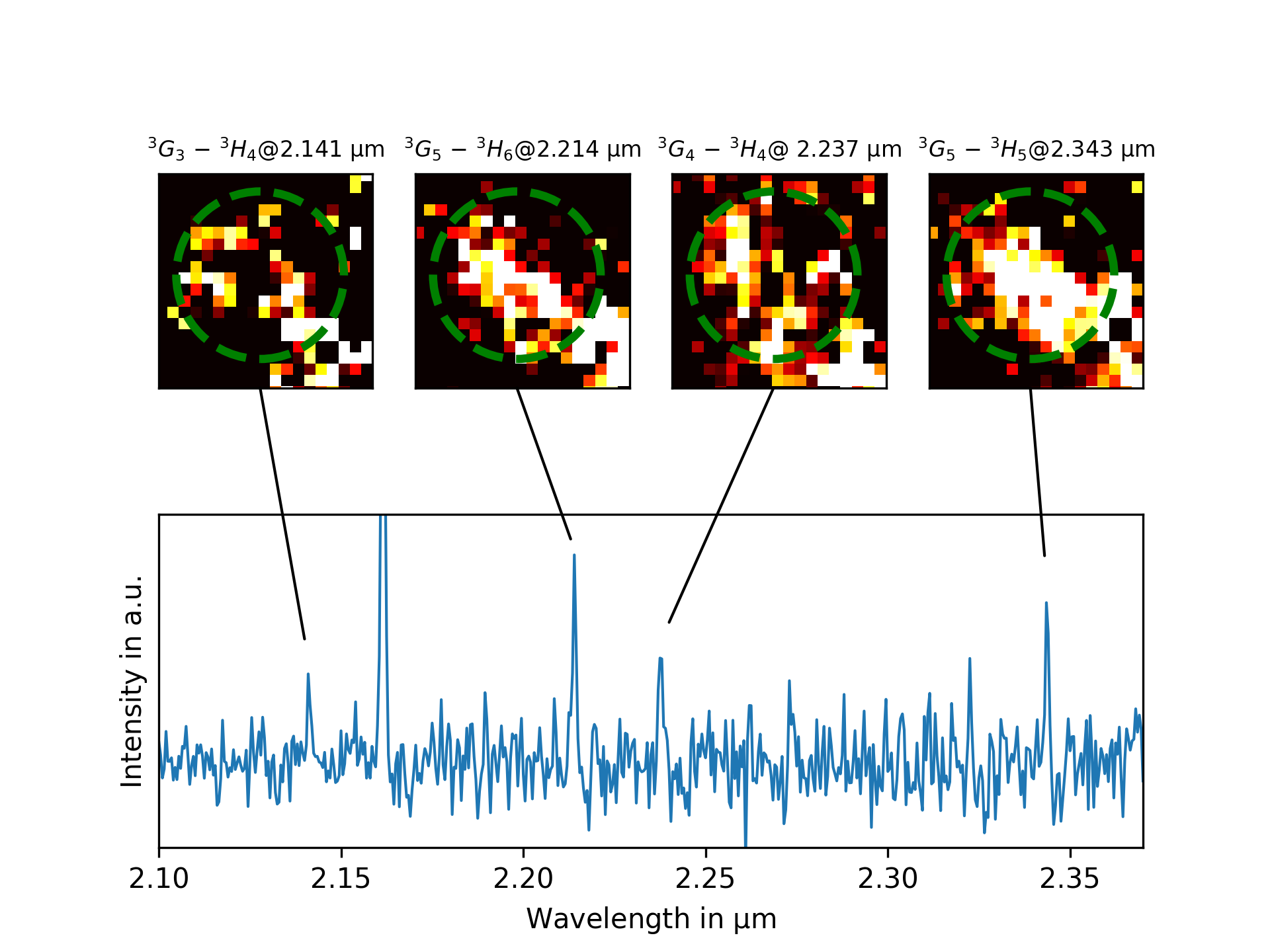}
	\caption{The spectrum of X7 in 2015. It shows the blue-shifted [Fe III] multiplet and the related $19\times 19$ px ($230\times 230$ mas) line-map cutouts. The prominent blue-shifted Br$\gamma$ can be found at $2.161\,\mu m$.}
	\label{X7_Fe_line}
\end{figure*}
Figure \ref{x8_line} shows the resulting line-maps when we select the mentioned red- and blue-wing of the emission line. The resulting line map that is achieved by selecting the $2.16455\,\mu m\,\, - \,\, 2.16516\,\mu m$ wavelength range is associated with the red lobe. It is slower than the blue lobe with respect to the Br$\gamma$ rest wavelength at $2.1661\,\,\mu m$. The corresponding blue lobe is obtained by selecting the $2.16394\,\mu m\,\, - \,\, 2.16455\,\mu m$ wavelength range. However, by selecting the whole Br$\gamma$ emission peak with a central wavelength of around $2.16455\,\,\mu m$, both lobes are visible in the line-maps that are presented in Fig. \ref{x8_line}. This excludes the possibility, that the lobes just represent the velocity of the source itself since the two emission peaks can be detected individually in 2006 and 2012. It should be noted that the distance between the faster blue lobe and Sgr A* is around $380\,\pm\,13$ mas. For the slower red lobe, it is $420\,\pm\,13$ mas. Hence, the faster blue lobe is closer to the SMBH as the slower red lobe.

\subsection{Continuum emission}

Because of the close distance of S41 (see Fig. \ref{2012_contour_map}), we are only able to provide an upper limit on the flux and brightness of X8.
The spectroscopic data of the emission area of X8 shows a weak H2 line that can be associated with the mentioned S-star. This is followed by the conclusion that X8 has to be fainter than S41. Figure \ref{2012_contour_map} shows the S-star S41 in 2012 that is located inside the magenta dashed circle. Figure \ref{2013_contour_zoom} shows the same situation but in 2013, where in contrast to the former image, X8 and S41 are disentangled. This shows that the PSF and therefore the data-quality influences the detection of X8. Sgr A* in Fig. \ref{2012_contour_map} and Fig. \ref{2013_contour_zoom} is located at the x (black and white). The contour lines in Fig.\ref{2012_contour_map} (left and right) are taken from the middle figure where we present a high-pass filtered version of the left K-band image. It shows the positional robustness of the high-pass filtering technique. A 3-pixel Gaussian is used to smooth the K-band image. After that, the smoothed output image is subtracted from the K-band image. The resulting smooth-subtracted image is again smoothed with a 3-pixel Gaussian. The final image shows the faint star S41 roughly at the position of X8. For a better comparison, a combination of the blue-shifted Br$\gamma$ line map of X8 and the continuum contour lines from the high-pass filtered image is shown in the right figure. 
For S41, we derive a magnitude of mag = $16.91\pm 0.06$ with a flux of $(0.107\pm 0.006)\times10^{-3}$Jy. The error is based on the literature values for the used calibration star S2 (see \citealt{schoedel2009}). In principal, the error should just cover the range between mag = $16.91+0.06$ because lower values would result in a brighter star. Considering that this S-star is not detected between 2006-2016, the K-band magnitude is most reasonable at the end of the range. 
In case X8 is as bright as S41, we would be able to detect an even brighter and elongated source since the flux would add up. From this, we can derive a lower K-band magnitude limit of mag = $16.97$ with a flux of $(0.107\pm 0.006)\times10^{-3}$Jy for X8. In Fig. \ref{2013_contour_zoom} we investigate the Galactic center with a low-pass filter. The zoomed area shows the S-stars S24/S25 and S41. X8 can be detected at the position of the blue-shifted Br$\gamma$ line map. Also remnants of X7 can be detected although the chance of confusion with S32 cannot be ruled out. The $90\%$ level contour lines of X8 are based on the blue-shifted Br$\gamma$ line map of 2013. Since the zoomed and low-pass filtered image is not convolved with a PSF because of the low iteration number, we are able to detect X8. Because of that, we conclude that the detected emission is produced by gas and dust.

\subsection{\bf [Fe III] line}

In addition to the presented $K_{S}$-band spectrum we report a simultaneously detected blue-shifted [Fe III] line for X7 as well as for X8. We identify the ionized blue-shifted [Fe III] multiplet ${}^3 G\,-\,{}^3 H$ for X7 and X8 in our data-set in 2015. Because of the limited FOV and the dynamic data quality, we emphasis the data-set of 2015 for comparing X7 and X8. The velocities, wavelengths, and the transition state of these lines can be found in Tab. \ref{tab2}. The mean velocity of the blue-shifted [Fe III] multiplet is slightly decreased compared to the mean velocity between the blue-shifted Br$\gamma$ and He$I$ line for both sources. The line-maps for X7 and X8 that are based on the blue-shifted [Fe III] lines and are in spatial agreement with the Br$\gamma$ and He$I$ analysis. These line-maps and the related spectrum can be found in Fig. \ref{X8_Fe_line} and Fig. \ref{X7_Fe_line}. Whilst the blue-shifted [Fe III] X8 detection is in agreement with the HeI and Br$\gamma$ emission, the X7 [Fe III] multiplet is dominated by the continuum for the  ${}^3 G_{3}\,-\,{}^3 H_{4} $ and ${}^3 G_{4}\,-\,{}^3 H_{4} $ lines. In contrast, the ${}^3 G_{5}\,-\,{}^3 H_{6}$ and ${}^3 G_{5}\,-\,{}^3 H_{5}$ lines coincide spatially with the blue-shifted HeI and Br$\gamma$ channel maps.
\begin{table*}[htbp!]
 								\centering
 								\begin{tabular}{cccccccc}
 									\hline\hline \\
 									\multicolumn{2}{c}{$[Fe III] line$} & \multicolumn{2}{c}{$X7$} & & \multicolumn{2}{c}{$X8$} & \\  
 									\\  \cline{1-2}\cline{4-5}\cline{7-8} Transition & Rest-wavelength  & & Blue-shifted line & Velocity & &  Blue-shifted line & Velocity   \\
 									\\ & $[\mu m]$ & &$[\mu m]$ & $[km\,s^{-1}]$ & & $[\mu m]$ & $[km\,s^{-1}]$ \\
 									 \hline
 									
 						                   	\\ ${}^3 G_{3}\,-\,{}^3 H_{4} $ & $2.1451$ & & $2.1410$ & $559$ & & $2.1441$ & $140$ \\
 							                   ${}^3 G_{5}\,-\,{}^3 H_{6} $ & $2.2178$ & & $2.2142$ & $488$ & & $2.2169$ & $122$ \\
 								               ${}^3 G_{4}\,-\,{}^3 H_{4} $ & $2.2420$ & & $2.2379$ & $549$ & & $2.2410$ & $133$ \\
 								               ${}^3 G_{5}\,-\,{}^3 H_{5} $ & $2.3479$ & & $2.3436$ & $549$ & & $2.3469$ & $128$ \\
 									\hline	\\
 								\end{tabular}
                                \caption[Fe III analysis]{The detected blue-shifted [Fe III] line velocities of X7 and X8 in 2015. The analytic uncertainty for determining the wavelength is $\pm\,2\,\times\,10^{-4}\,\mu m$, the velocity error therefore $\pm\,20\,km\,s^{-1}$. The mean blue-shifted [Fe III] velocity for X8 is around $130,km\,s^{-1}$. For X7, it is $536\,km\,s^{-1}$.}						
                                \label{tab2}
\end{table*}

\section{Discussion}
\label{secDis}

In this section, we discuss several possible scenarios concerning the nature of X8.

\subsection{X8 detection and the optical depth}

Based on the detection presented in Fig. \ref{2013_contour_zoom}, we can confirm a continuum counterpart to the blue-shifted line detection of X8. 
The detected source in the continuum shows a similar elongated shape compared to its gaseous counterpart. The elongation is underlined by the $x$- and $y$-expansion measurements presented in Table \ref{tab1}. In both directions, the dimension of X8 exceeds the SINFONI PSF. Therefore, the observed source emission is extended and not compact. We detect two separate main emission peaks. These two peaks can be each linked to the detected lobes. This could indicate a more complex structure. Because of the faster lobe that is located closer to Sgr A* than the slower lobe, it can be speculated that the presence of a possible wind from Sgr A* or a wind of the source itself causes this setup. Since the proper motion is directed away from the SMBH, this asymmetrical system could be influenced by either external or internal winds. Another possibility could be an outflow from massive stars around Sgr A*. \cite{muzic2007} concluded that a wind from massive stars around Sgr A* is a less plausible scenario. When it comes to the optical depth, the $\frac{HeI}{Br\gamma}$ ratio is $\sim 0.7$. This indicates optically thin material (see also \citealt{shcherbakov}). Even though the continuum detection could be partially dominated by the blue-shifted line emission, the continuum counterpart of X8 can be assumed to be optically thick stellar or dust photosphere. To distinguish between potential stellar and dust contributions to the continuum, further observations in L'- and M-bands are needed.

\subsection{The closest bipolar outflow source to Sgr A*}

As shown by \citet{Yusef-Zadeh2017}, eleven bipolar-outflow sources are close to Sgr A* and are located within a projected distance of 1 pc from the SMBH. The authors report sources that show similarities to the detected features of X8. The detected source BP1 \citep{Yusef-Zadeh2017} is described as two bright lobes that are embedded in a molecular cloud. This scenario could explain the detection of X8 that we present in Fig.\ref{x8_stacked}. Also, the detection of GL2591 (see \cite{GL2591} for more details) show a similar 2D cut as we find for X8 (Fig. \ref{x8_stacked}). To verify the detection of the two-lobe feature, we also show Br$\gamma$ line-maps, where we selected just the red- and blue-shifted shifted wing of the Doppler-shifted Br$\gamma$ emission peak. As shown in Fig.\ref{x8_line}, the detection of the two lobes depends on the selected wing of the Br$\gamma$ emission peak. This can be translated to the identification of a blue- and red-shifted lobe for X8 whereas the blue lobe is closer to Sgr A*. Hence, the red lobe is further away from the super massive black hole. Since the quality of the data is not stable due to changing weather conditions, we present two examples as well as a stacked version of the two data-sets. Even though noise and background features in the presented data increase the level of confusion for the detection of X8, the results presented in Fig.\ref{x8_line} clearly underlines a two-lobe scenario.  

\subsection{X8 - a possible bow-shock source}

Apart from exhibiting a bipolar morphology, the outflow associated with X8 is also likely to drive a shock into the surrounding medium as most stars in the inner parsec are expected to move close to the local sound speed \citep[see][especially their Fig. 1]{2017bhns.work..237Z}. The $Br\gamma$ emission of X8 tracing ionized gas appears to be directed towards Sgr A*. The position angle between Sgr A* and X8 is around 45° in 2006 and 55° in 2016. Since we detect a proper motion of X8, the change in angle can be explained by its trajectory. Especially in Fig. \ref{finding_chart}, which is based on the data-set in 2015, shows similarities between the bow-shock source X7 and X8. However, we also observe the influence on the data that is caused by the background and overall quality (see Fig. \ref{2012_contour_map} and \ref{2013_contour_zoom}). It is therefore a reasonable possibility that the two-lobe structure of X8 in 2015 cannot be detected because of the mentioned influence. Under the assumption that the X8 drives a bow shock into the ambient medium, the bipolar-outflow of X8 could be influenced by a potential external wind from the direction of Sgr A* (see \citealt{muzic2010}). It would be the third bow-shock source that (at least in projection) lies in the mini-cavity, which is supposed to be generated by the mentioned outflow. The detected blue-shifted [Fe III] line could also be a possible indicator for wind that spatially originates at the position of SgrA* since the ionization of iron needs high energies (around 35000K, see \cite{likkel2006} for further information). The line-ratio between [Fe III] and Br$\gamma$ for planetary nebulas (PNs), which is found in the mentioned publication, is around $2\%$. However, the line-ratio between [Fe III] and Br$\gamma$ for X8 is around $45\%$. Therefore, we can exclude the intrinsic  scenario and the external origin for [Fe III], in particular a bow shock, are more likely. In other words, if the excitation cannot be explained completely by intrinsic processes, an external wind already inferred from the detected X3 and X7 sources \citep{muzic2010} can explain such a high line ratio. In addition, the presence of an outflow also determines the overall bow-shock orientation if the outflow velocity is larger than the stellar velocity since the bow-shock is oriented along the relative velocity, $\mathbf{v}_{\rm rel}=\mathbf{v}_{\star}-\mathbf{v}_{\rm ambient}$ \citep[see also the discussion in][]{muzic2010,Zajacek2016}. The fact that we find comparable [Fe III] to Br$\gamma$ ratios for X7 of around $40\%$ could point to a similar nature of X8 and X7.

\subsection{Orbital constraints, location within NSC, and basic hydrodynamics of X8}

The X8 source with the mean projected distance $R\sim 425 \pm 26\,{\rm mas}=0.017\pm 0.001\,{\rm pc}$ appears to lie in the innermost parts of the NSC in either the clock-wise disc or S-cluster, which are inside $\sim 0.4\,{\rm pc}$. If confirmed, the presence of either a protostar or an AGB star would have implications for their origin and dynamics since previously it was assumed that S-cluster members are rather B-type stars/dwarfs (spectral type in the range of B0-B3V) with an upper age limit of 15 Myr \citep{2017ApJ...847..120H} and the clock-wise disc stars are of spectral type OB with the similar age constraint of a few million years \citep{2010ApJ...708..834B}.
\begin{figure}[h!]
    \centering
    \includegraphics[width=0.5\textwidth]{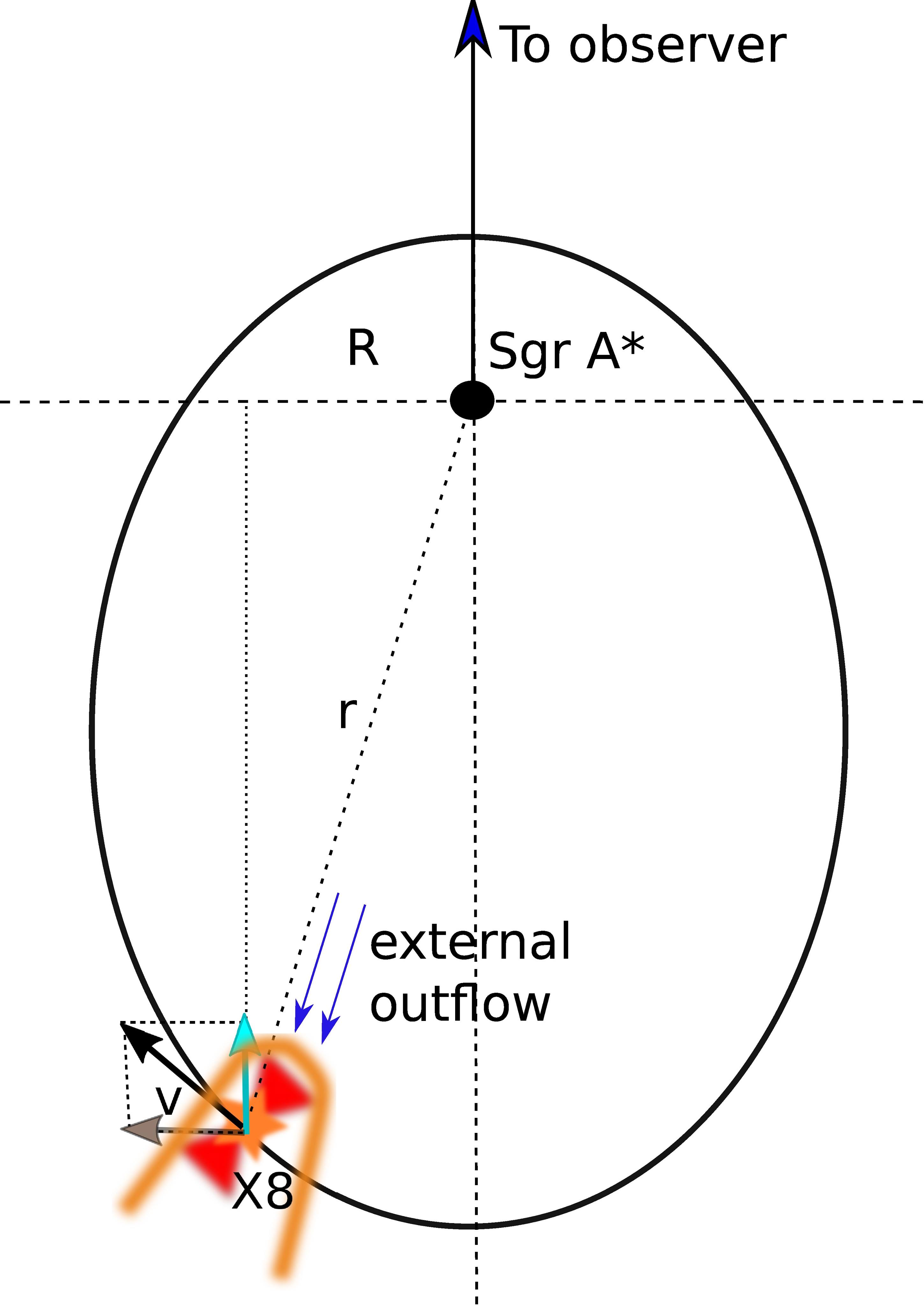}
    \caption{Illustration of the eccentric X8 and its current position in the background with respect to Sgr~A*. Bipolar morphology and the bow-shock orientation are also depicted.}
    \label{fig_orbitX8_illustration}
\end{figure}
Since X8 is well inside the gravitational influence of Sgr~A*, which is estimated for the Galactic centre to be $r_{\rm h}\approx 2\,{\rm pc}$, we can assume that it is bound and orbits Sgr~A* on an elliptical orbit as other S stars. Its radial velocity of $v_{\rm r}=215\,{\rm km\,s^{-1}}$ is blue-shifted, whereas the tangential component $v_{\rm t}=(293\pm 20)\,{\rm km\,s^{-1}}$ points away from Sgr~A*. The total space velocity of the star then is $v_{\star}=(v_{\rm r}^2+v_{\rm t}^2)^{1/2}=363.4\,{\rm km\,s^{-1}}$. Since the radial and the tangential velocity components are of a comparable magnitude, we can assume that the X8 orbit is close to edge-on with the current X8 position in the background with respect to Sgr~A* and moving towards us, see Fig.~\ref{fig_orbitX8_illustration} for an illustration. We can also infer the position along the orbit - the true anomaly $\phi=90^{\circ}+\phi_0$ from the decomposition of the velocity vector, $\tan{\phi_0}=v_{\rm t}/v_{\rm r}$. from which $\phi=90^{\circ}+53.7^{\circ}=143.7^{\circ}$.

The orbit is not consistent with zero eccentricity. In case the orbit was circular, we could estimate the 3D distance from Sgr~A* simply as $r\approx GM_{\bullet}/v_{\star}^2=0.13\,{\rm pc}$. From Fig.~\ref{fig_orbitX8_illustration}, the orbital phase then would be $\phi'=172.8^{\circ}$, which is in discrepancy with the velocity-based calculation. Instead, we can look for a series of orbital solutions $(a,e)$ with a non-zero eccentricity $e$ and semi-major axis $a$ given the true anomaly of $\phi=143.7^{\circ}$ and the total velocity $v_{\star}=363.4\,{\rm km\,s^{-1}}$. An exemplary calculation can be done for the mean eccentricity $\overline{e}=2/3$ of the thermal distribution of eccentricities $n(e)\mathrm{d}e=2e\mathrm{d}e$, which appears to be a good approximation for the S stars \citep{2010RvMP...82.3121G}. By combining the velocity and radius equations for the elliptical orbit, we obtain the semi-major axis of X8, $a_{X8}=0.09\,{\rm pc}$. The position of X8 with respect to the observer is illustrated in Fig.~\ref{fig_orbitX8_illustration} for a moderately eccentric orbit.
\begin{figure}[h!]
    \centering
    \includegraphics[width=0.5\textwidth]{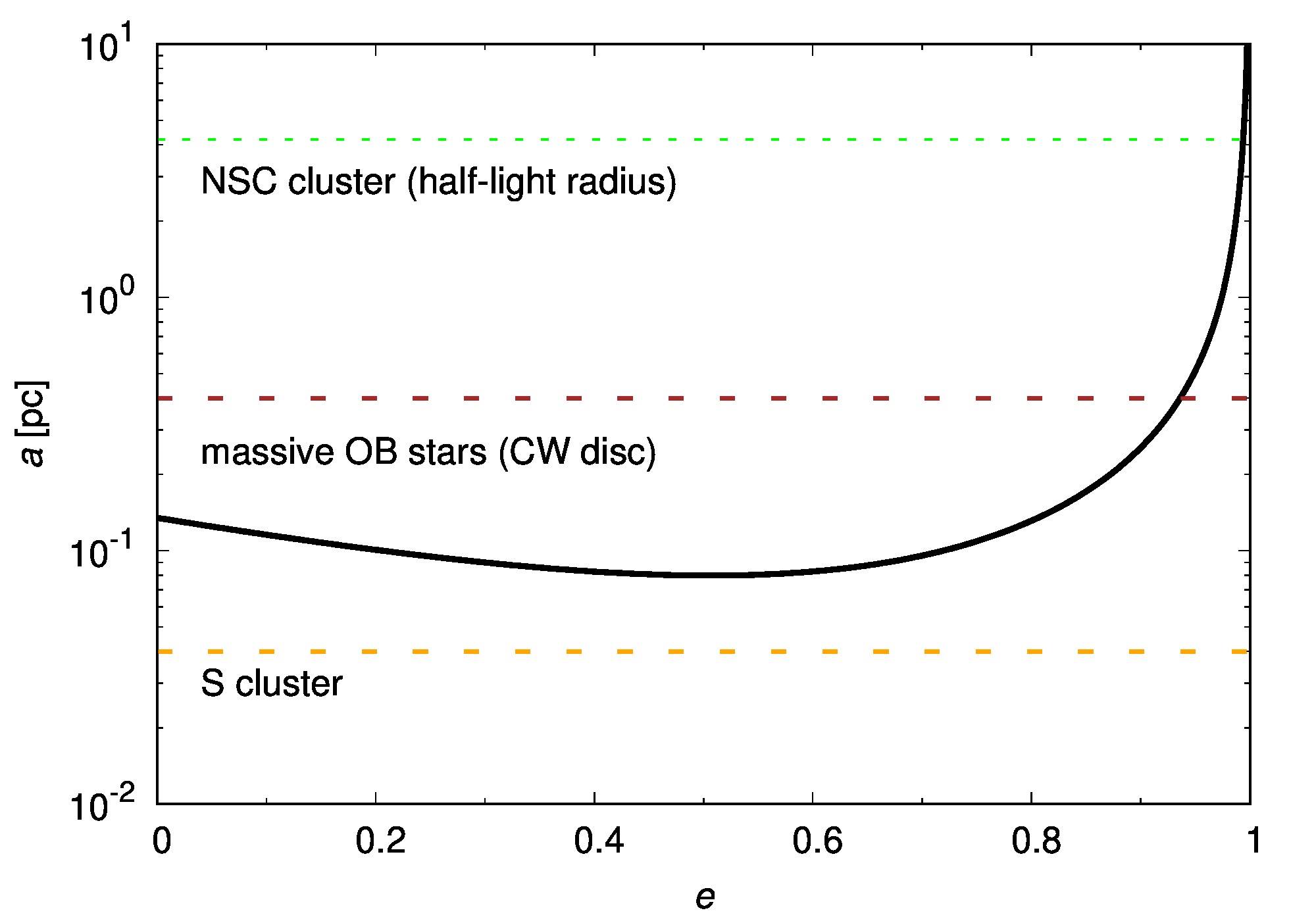}
    \caption{Calculated semi-major axis of the X8 source as a function of its eccentricity assuming an edge-on elliptical orbit. The horizontal lines mark the characteristic scales of the known young stellar populations in the Galactic center region: S stars and CW disc members \citep{2010RvMP...82.3121G}. The uppermost line marks the half-light radius of the NSC, $r_{\rm hl}=(4.2 \pm 0.4)\,{\rm pc}$, according to \citet{2014CQGra..31x4007S}.}
    \label{fig_a_e_x8}
\end{figure}
The semi-major axis of X8 as a function of eccentricity in the range $(0,1)$ is plotted in Fig~\ref{fig_a_e_x8}. The semi-major axis is of the order of $a_{\rm X8}\approx 0.1\,{\rm pc}$ since the eccentricity values larger than $0.9$ are unlikely. Hence, X8 falls into the same range of semi-major axes as the young massive stars in the clock-wise disc and the scattered population \citep{2010RvMP...82.3121G}. Although corrections to these estimates are necessary with a proper orbital analysis with more data, we have at least shown that the X8 source is fully consistent with being located inside the NSC in its inner parts.

The X8 moves away from Sgr~A* but the symmetry axis of Br$\gamma$ extended emission points towards Sgr~A*. This can be explained by the bow-shock formation when the orbiting X8 interacts with an outflow of $v_{\rm out}\approx 1000\,{\rm km\,s^{-1}}$ from the direction of Sgr~A*, in a similar way as for the comet-shaped sources X3 and X7 \citep{muzic2010}. The characteristic size of the bow shock is given by the stagnation radius \citep{1997ApJ...474..719Z},
\begin{equation}
     R_0=\left(\frac{\dot{m}_{\rm w}v_{\rm w}}{\Omega \rho_a v_{\rm rel}^2} \right)^{1/2}\,,
     \label{eq_stagnation_radius}
\end{equation}
at which the stellar wind pressure and the ram pressure are at equilibrium. In Eq.~\eqref{eq_stagnation_radius} $\dot{m}_{\rm w}$ and $v_{\rm w}$ are the stellar mass-loss rate and the terminal wind velocity, respectively, $\Omega$ is the solid angle into which the stellar wind is blown. $\rho_{\rm a}$ denotes the ambient density and $v_{\rm rel}$ is the relative velocity of X8 with respect to the outflowing ambient medium. The solid angle is given by $\Omega=2\pi(1-\cos{\theta_0})$, where $\theta_0$ is the opening angle of the outflow.
\begin{figure}[h!]
    \centering
    \includegraphics[width=0.5\textwidth]{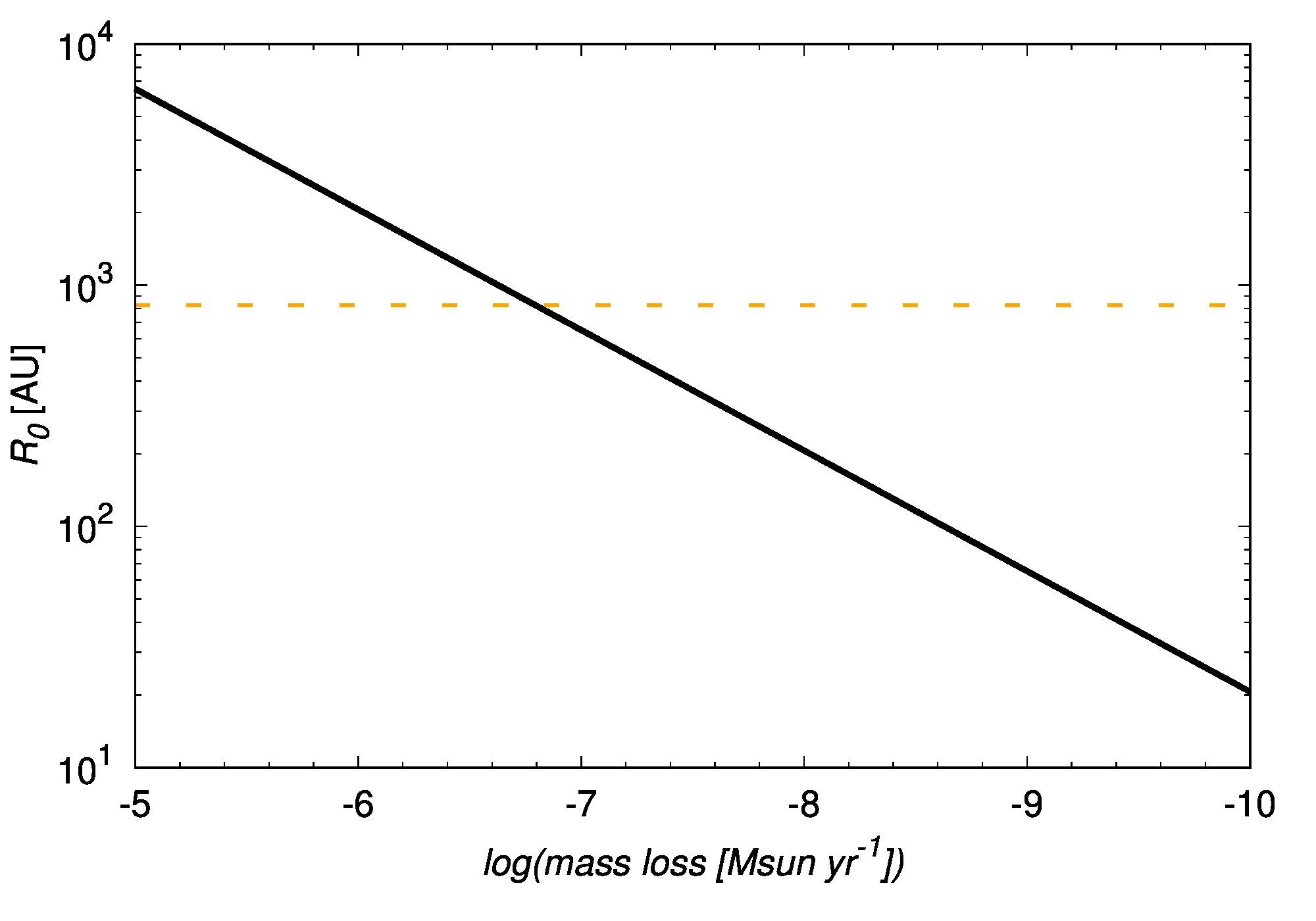}
    \caption{The stagnation radius of a stellar bow shock as a function of its mass-loss rate according to Eq.~\eqref{eq_stagnation_radius}. The other parameters were fixed based on observed values as discussed in the text. The horizontal line marks the stagnation radius as estimated from the brightness profile in Fig.~\ref{x8_stacked}.}
    \label{fig_stagnation_radius}
\end{figure}
For X8 we can estimate several parameters in Eq.~\eqref{eq_stagnation_radius}. The stagnation radius of X8 is about $R_0\sim 100\,{\rm mas}\sim 825\,{\rm AU}$ as indicated by the intensity profile in Fig.~\ref{x8_stacked}. The terminal wind velocity can be estimated based on the FWHM of Br$\gamma$ line, $v_{\rm w}\sim 170\,{\rm km\,s^{-1}}$. The opening angle of the bipolar outflow can be approximated by $\theta_0\sim 45^{\circ}$ again from the brightness distribution in Fig.~\ref{x8_stacked}. The relative velocity vector is given by the vector difference between the stellar motion and ambient medium flow. With the approximation that the external outflow is directed perpendicular to the orbital velocity, the relative velocity magnitude can be estimated as $v_{\rm rel}\approx \sqrt{v_{\star}^2+v_{\rm out}^2}\sim 1064\,{\rm km\,s^{-1}}$. Finally, the ambient density at the estimated deprojected distance of X8, $r_{\rm X8}\sim 0.1\,{\rm pc}$, can be calculated from the ADAF profile of \citet{2006ApJ...640..319X},  $n_{\rm a}\sim 10^{4}(r/10^{15}\,\rm cm)^{-1}\,{\rm cm^{-3}}$, with the mass density $\rho_{\rm a}\approx m_{\rm H}n_{\rm a}$. Subsequently, this allows us to search for the mass-loss rate of X8 that is consistent with the observed bow-shock size of $R_0$. In Fig.~\ref{fig_stagnation_radius}, we plot the stagnation radius $R_0$ as a function of the mass-loss rate. We see that the observed stagnation-point size of $100\,{\rm mas}$ is consistent with the mass-loss rate of X8 of $\dot{m}_{\rm w}=10^{-6.8}\,M_{\odot}{\rm yr^{-1}}$, which is a realistic value for both protostars and post-AGB stars as we discuss further in the following subsection.

\subsection{Plausible scenarios for the origin of X8}

The elongated, bipolar geometry of X8 is consistent with two scenarios that we discuss below:
\begin{itemize}
 \item \textit{Dust-embedded young stellar object (YSO)}. In general, young stellar objects are naturally characterized by axisymmetric geometry due to the presence of rotationally flattened dusty envelopes \citep{1976ApJ...210..377U}. Depending on the YSO stage \citep[stage 0, I, II or III according to][]{1987IAUS..115....1L}, the envelope is progressively getting geometrically thinner and accretion rate onto the star becomes smaller. The X8 source would be consistent with YSO stage of class I surrounded by optically thick dusty envelope dissected by bipolar outflows -- such a scenario was also envisaged for the unresolved Dusty S-cluster Object (DSO/G2), see \citet{Shahzamanian2016} and \citet{Zajacek2017} for detailed radiative transfer simulations. In a similar way, the massive young stellar object NGC 3603 IRS 9A* shows clear signs of the disc--bipolar cavity presence \citep{2016A&A...588A.117S}. The Br$\gamma$ lobes could be associated with the ionized outflows present in the cavities. The bipolar outflow is expected to further interact with the ionized wind from the direction of Sgr~A* \citep{muzic2010,2014IAUS..303..150S}, with the estimated terminal velocity of $\sim 1000\,{\rm km\,s^{-1}}$, which is expected to result in a bow shock driven into the ambient medium \citep{2013ApJ...768..108S}. This shock would be rather diluted, but could contribute to the overall elongation of X8  along the Sgr~A*--X8 connection line. The dusty envelope would be oriented in a perpendicular direction to the outflow, see the illustration in Fig.~\ref{x8_sketch} for the geometrical illustration of X8 as an YSO source.
  \begin{figure}[htbp]
	\centering
	\includegraphics[width=0.5\textwidth]{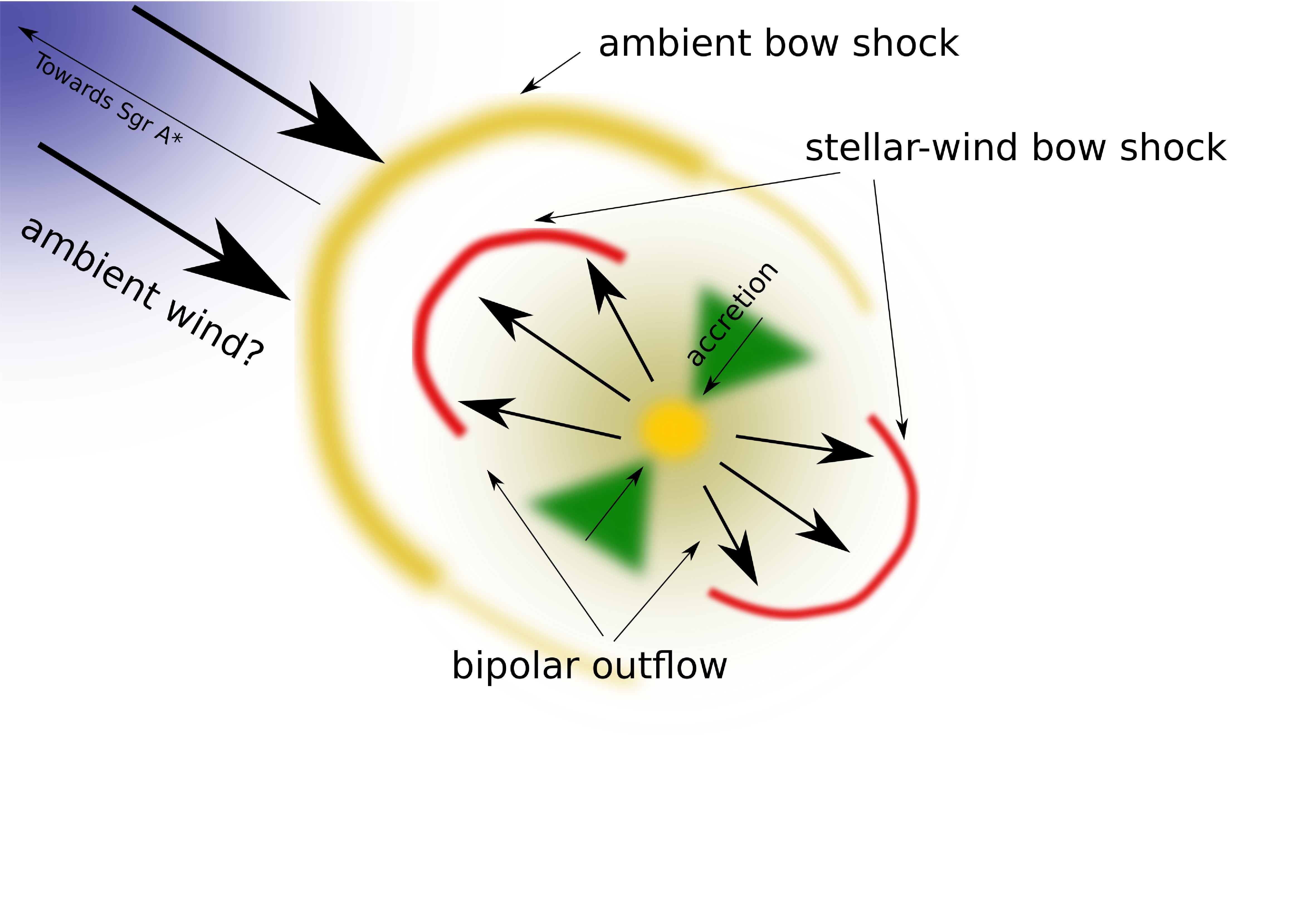}
	\caption{Illustration of the possible interpretation of the X8 object in the context of a bipolar young stellar object of class I. We show the approximately 0".1 (4 mpc) diameter region around the source X8. The optically thick dusty envelope is intersected by bipolar outflow that forms cavities. The X8 may further interact with the ambient ionized outflow from the direction of Sgr~A*, driving a shock into the ISM on the side facing Sgr~A*. The accretion of material onto X8 would then proceed in the direction perpendicular to the bipolar outflow.}
	\label{x8_sketch}
\end{figure}
  \item \textit{Post-AGB star/Young planetary nebula.} Apart from young stellar objects, bipolar outflows are associated with the final stages of the stellar evolution, namely with collimated winds from red giants and post-AGB stars in the early planetary nebula phase \citep{1999NewAR..43...31F}. The bipolarity could be formed during the early planetary nebula stage, when the detached gas from aging stars is collimated by magnetic field or by a companion star \citep{2000oepn.book.....K,2017MNRAS.468.4465C}. Bipolar planetary nebulae resemble young stellar objects in many characteristics, namely emission tracing ionized gas with comparable line widths.
\end{itemize}
Without further spectroscopic evidence, one cannot clasify X8 definitely as either a YSO or a post-AGB star. Statistically, given the presence of other young stars in the region, including proplyd, bipolar, and dusty sources \citep{EckartAA2013,2015ApJ...801L..26Y,Yusef-Zadeh2017}, one can hypothesize that X8 is another manifestation of a recent star-formation event. However, both the YSO and young planetary nebula scenarios imply that X8 belongs to the extreme populations of the nuclear star cluster -- to either very young or old stars and out of the main sequence. More detailed analysis of this and similar sources can complement previous population synthesis studies \citep{2009A&A...499..483B,2018A&A...609A..26G} and thus shed more light on the rich episodic star-formation history of the nuclear star cluster \citep{2011ApJ...741..108P}. 

Given the observational properties of X8 and its stability over the years, we consider the following scenarios, which could in theory resemble a bow-shock as well as bipolar morphology, unlikely, 
\begin{itemize}
  \item \textit{Supernova}. Massive stars present in the central parsec of the Galactic centre will end their lives as supernovae with the approximate rate of $\dot{M}_{\rm SN}\approx 10^{-4}\,{\rm yr^{-1}}$. The supernova explosions are not symmetric, which results in initial neutron and black hole velocity kicks. The initial expansion velocities of the supernova shell are, however, one order larger than what we measure for the Br$\gamma$ line, of the order of $\sim 1000\,km\,s^{-1}$. Moreover, supernova explosions would manifest themselves by non-thermal radio emission and  high-energy radiation -- in X-rays and $\gamma$-rays. Since no such an event has been detected recently, we consider the supernova origin of X8 unlikely,
  \item \textit{Core-less cloud}. The overall stability and bipolar nature of the source are inconsistent with a gaseous-dusty cloud without any stellar core. Any core-less cloud would have a tendency to get dispersed in a highly ionizing medium and/or become more elongated/filamentary due to the interaction with the nuclear outflow. This can be underlined by the presence of the detected blue-shifted [Fe III] multiplet ${}^3 G\,-\,{}^3 H$. 
\end{itemize}

\section{Conclusion}
\label{secCon}
The aim of this analysis can be summarized as an overview of possible scenarios for the object X8. The fact, that we observe X8 in every year in the blue-shifted emission line-maps in combination with the K-band continuum detection points to a star-like scenario with a possible bow-shock. Also, the detected two lobes show different velocities. The lobe closer to Sgr A* is faster than the lobe, that is further away. This could be explained with intrinsic properties. But also, external effects that cause this velocity gradient cannot be ruled out. Since this setup of two different lobes with different velocities can be observed over a time-span of almost 15 years, a central star that stabilizes this system is a reasonable explanation. In contrast, a pure gas-cloud identity close to the S-stars S24, S25, and S41 would result in dissolving scenarios. Also, gas would probably not be arranged in a complex structure with two individual detectable lobes that show different velocities. The detected blue-shifted [Fe III] can be associated with high temperatures that also neglect a pure gaseous nature of X8. If future observations can reveal the stellar component of X8, the source would be indeed one of the closest YSO to Sgr A* to date. Since X8 is moving away from S24 and S25, the object becomes more isolated. A more detailed analysis that is covered by NIR observations could provide more properties of the two lobes. These findings could then confirm the closest bipolar outflow source to Sgr A*. But already today we observe strong evidence that underline this possibility. The two individual lobes can be observed in our high-quality stacked images as well as in the data-cubes of different years. The limiting factor is in every case the quality and amount of observations that cover the western region next to S24/S25. Nevertheless, this indicates that the two-lobe feature is not a stacking feature. The individual lobes, that are shown in the line-maps, are extracted by selecting the blue- and red-lobe of the blue-shifted Br$\gamma$ peak (with respect to the rest-wavelength) and underlines therefore the presence of two individual lobes with two velocities.  
	
\section*{Acknowledgements}We highly appreciate the comments of the anonymous referee that helped to improve this paper.
We received funding
from the European Union Seventh Framework Program
(FP7/2013-2017) under grant agreement no 312789 - Strong
gravity: Probing Strong Gravity by Black Holes Across the
Range of Masses. Michal Zaja\v{c}ek acknowledges the support by the National
Science Centre, Poland, grant No. 2017/26/A/ST9/00756 (Maestro 9).
This work was supported in part by the
Deutsche Forschungsgemeinschaft (DFG) via the Cologne
Bonn Graduate School (BCGS), the Max Planck Society
through the International Max Planck Research School
(IMPRS) for Astronomy and Astrophysics, as well as special
funds through the University of Cologne and
SFB 956 – Conditions and Impact of Star Formation. Part of this
work was supported by fruitful discussions with members of
the European Union funded COST Action MP0905: Black
Holes in a Violent Universe and the Czech Science Foundation
- DFG collaboration (No. 13-00070J). We also would like to thank the members of the SINFONI and ESO Paranal/Chile team.

\bibliographystyle{aa}
\bibliography{output.bbl}
\end{document}